\documentclass[aps,preprint,amssymb,12pt,floatfix]{revtex4}
\usepackage{subfigure}
\usepackage{graphicx}
\usepackage{rotating} 
\usepackage{color}
 \usepackage{amsmath,amsthm}
\usepackage{epstopdf}
\usepackage{wrapfig}

\begin{document}
\title{Development and applications of Coarse Grained models for RNA}
\author{Changbong Hyeon}
\affiliation{School of Computational Sciences, Korea Institute for Advanced Study, Seoul 130-722, Republic of Korea}
\author{Natalia A. Denesyuk and D. Thirumalai}
\affiliation{Biophysics Program, Institute For Physical Science and Technology, University of Maryland, College Park, MD 20742, USA}
\date{\today}
\begin{abstract}
In contrast to proteins much less attention has been focused on development of computational models for describing RNA molecules, which are being recognized 
as playing key roles in many cellular functions. Current atomically detailed force fields are not accurate enough to capture the properties of even simple nucleic acid 
constructs. In this article, we review our efforts to develop coarse-grained (CG) models  that capture the underlying physics for the particular length scale of interest. 
Two models are discussed. One of them is the Three Interaction Site (TIS) model in which each nucleotide is represented by three beads corresponding to sugar, phosphate, 
and  base. The other is the Self-Organized Polymer (SOP) model in which each nucleotide is represented as a single interaction center. Applications of the TIS model to study the complexity of hairpin formation and the effects of crowding in shifting equilibrium between two conformations in human telomerase pseudoknot are 
described. The work on crowding illustrates a direct link to the activity of telomerase. We use the SOP model to describe the response of {\it Tetrahymena} ribozyme to 
force. The simulated unfolding pathways agree well with single molecule pulling experiments. We also review predictions for the unfolding pathways for {\it Azoarcus} 
ribozyme. The success of the CG applications to describe dynamics in RNA gives hope that more complex processes involving RNA-protein interactions can be tackled using 
variants of the proposed models.
\end{abstract}
\maketitle

The use of models in obtaining insights into the organization of nucleic acids can be traced back to the publication in 1969 of a seminal paper \cite{Levitt69Nature} 
outlining a detailed structure for transfer nucleic acid (tRNA) by Levitt, one of the recipients of the 2013 Nobel prize in chemistry. By assuming that
conformations of different tRNA molecules are similar (homologous) and the association of hydrophobic moieties stabilized by hydrogen bond 
interactions and hydration of charged and other solvent exposed polar groups are free energetically favorable, Levitt predicted with stunning accuracy a detailed 
structure of tRNA. Although there has not been a great deal of attention paid to RNA structures until recently,  several major discoveries have made it abundantly clear 
that  RNA is not merely a passive carrier of information but is involved in many cellular functions (many are most likely in the category of ``unknown unknowns").  Thus,  
it is important not only to determine RNA structures but also how they fold (the RNA folding problem) \cite{Tinoco99JMB}, and function in the crowded cellular 
environments, and respond   to binding of metabolites as occurs in the process of gene expression in  bacteria. The growing importance of RNA has ushered this important 
class of molecules at the center stage of biology, raising questions that were once familiar only in the study of proteins.  From a computational perspective of RNA folding,
the dream would be to generate movies using atomically detailed molecular dynamics simulations starting from different initial conditions till  RNA reaches the folded 
structure. 
For RNA this is not yet a reality because force fields are not accurate \cite{Garcia13PNAS} and the folding time scales are extremely long (for large ribozymes it can exceed seconds). 
This forces us to use concepts  rooted in polyelectrolyte theory and coarse-grained (CG) models to make much needed progress, and in the process discover general principles
of RNA folding. Here, we describe some developments along these lines that we have made over the past decade or more.  

The use of CG models, based on theoretical ideas, to describe the essential phenomena in a number of areas has a rich history in science. Insights into many problems in 
condensed matter physics have come from the use of minimal CG models that capture the essence of complex phenomena. Such models  can sometimes be solved by purely 
analytical methods. However, complex problems, such as spin glasses \cite{Mezardbook},  structural glasses \cite{Kirkpatrick89JPhysA,Parisi2010RMP,Biroli2013JCP}, and a 
host of problems in biology such as RNA folding \cite{Thirum05Biochem}, and enzyme functions \cite{Warshel76JMB} require computational methods covering a broad spectrum of 
time and length scales.  With increasing computational power and continued focus on developing increasingly accurate force-fields, in the spirit of pioneering works in the 
context of protein folding and dynamics \cite{Levitt76JMB,Levitt75Nature,McCammon77Nature} and enzyme reactions, it is likely that in the not too distant future one will be 
able to represent the folding of proteins in atomic detail. In fact, rapid progress has been made in obtaining folding trajectories of a few single domain 
proteins \cite{Shaw10Science,Lindorff2011Science,Lane2012COSB}.  
At present, the prospects of performing atomically detailed and accurate molecular dynamics (MD) simulations for nucleic acids seems daunting. Given the paucity of 
predictions for RNA folding using MD simulations, there should be interest in developing coarse-grained (CG) models for nucleic acids with the view towards describing many 
aspects of self-assembly of RNA folding, folding of chromatin, and protein-RNA (and DNA) interactions. Although there are several examples of using simulations of 
coarse-grained models in condensed matter and the study of proteins, there is still a good deal of resistance to the use of CG models to make testable 
predictions in biology. 
The purpose of this article is to document a few case studies from our own research to  illustrate the predictive power of CG models in the difficult area of RNA folding. 

Modeling reality requires a level of abstraction, depending on the phenomenon of interest. For example, near a critical point of a fluid (or a magnet), exponents that 
describe the vanishing of the order parameter (density or magnetization) or divergence of the correlation length (describing density-density  or spin-spin correlation 
function) are universal, depending only on the spatial dimensionality ($d$), and are impervious to  atomic details.  
These findings, which are rooted in the concepts of universality and renormalization group \cite{Fisher74RMP},  are also applicable to the properties of 
polymers \cite{deGennesbook}. In nucleic acids, at short length scales ($l\lesssim 5$ \AA) detailed chemical environment determines the basic forces (hydrogen bonds and dispersion forces) between two nucleotides. 
If $l\sim1$--1.5 nm, interactions between two bases, base stacks and grooves of the nucleic acids become relevant. Understanding how RNA folds 
($l\sim$1--3 nm) requires 
energy functions that provide at least a CG description of nucleotides and of interactions between them in the native state and excitations
around the folded structure. On the persistence length scale $l_p\approx 150$ bp $\approx 50$ nm \cite{Bustamante94SCI} and beyond it suffices to treat dsDNA as a stiff elastic filament without explicitly capturing the base-pairs.  
If $l\sim \mathcal{O}(1) $ $\mu$m dsDNA behaves like a self-avoiding polymer \cite{VallePRL05}. 
On the scale of chromosomes ($l\sim$ mm) a much coarser description suffices. 
Thus, models for DNA and RNA vary because the scale of structural organization changes from nearly millimeters in chromatin to a few nanometers in the folded states of RNA. 
In order to describe RNA folding and the response of these molecules to mechanical forces we have introduced two classes of CG models. The first represents each nucleotide by three interaction sites, referred to as the TIS model.  Here, we employ the TIS model to illustrate the complexity of 
RNA hairpin formation and the effects of crowding on the equilibrium shift between pseudoknot and hairpin in human telomerase RNA. In the second model, 
each nucleotide is represented using a single interaction center. The efficacy of the resulting model, referred to as the Self-Organized Polymer (SOP) model, is illustrated
using simulations to predict the outcomes of single molecule pulling experiments. 

\section{General considerations in modeling RNA}

In general, as a natural consequence of rugged folding landscape, folding of RNA is partitioned into several pathways, the effect of which can be detected  in the form of multi-exponential functions in kinetics experiments \cite{Thirum05Biochem}
For several reasons the folding landscape of RNA is rugged consisting of multiple basins of attractions separated by substantial free energy barriers.   
(i) Phosphate groups are negatively charged, which implies that polyelectrolyte (PE) effects  oppose collapse and folding of RNA. 
Valence, size and shape of counterions, necessary to induce compaction and folding \cite{ThirumARPC01}, can dramatically alter the thermodynamics and kinetics of RNA folding. Thus, PE effects, which are controlled by a number of factors play a key role in determining RNA folding thermodynamics  and kinetics (see below).
(ii) The purine and pyrimidine bases of nucleotides have different sizes but are chemically similar. 
(iii) Only $\sim 54$ \% of bases form canonical Watson-Crick base pairs while the 
remaining nucleotides are in non-pairing regions \cite{Dima05JMB} (discussed  further below). 
(iv) The lack of chemical diversity in the bases results in RNA easily adopting alternate stable misfolded conformations, resulting in a relatively small stability gap between the folded and misfolded structures \cite{Thirum05Biochem}. 
Thus, the homopolymer nature of the RNA monomers, the critical role of counterions in shaping the folding landscape, and the presence of multiple low-energy excitations around the folded state make RNA folding a challenging problem. \\

\subsection*{Polyelectrolyte (PE) effects}
To fold, RNA must overcome the large electrostatic repulsion between the negatively charged phosphate groups. 
PE based theory (see Box.\ref{PE}) shows that multivalent cations ($Z>1$) are more efficient in neutralizing the backbone charges than monovalent ions --
a prediction that is borne out in experiments. For example, the midpoint of the folding transition $C_m$, the ion-concentration at which the populations of
the folded and unfolded states are equal, for \emph{Tetrahymena} ribozyme is $\sim 3\times 10^6$ fold greater in Na$^+$ than in cobalt-hexamine ($Z=3$)! 
The nature of the compact structures depends on $Z$, with the radius of gyration scaling as  $R_G\propto 1/Z^2$ \cite{WoodsonJMBI01}, implying compact intermediates have lower free energy as $Z$ increases.  
Thus, folding rates should decrease as $Z$ increases, which also accords well with experiments \cite{moghaddam09JMB}. 
Polyelectrolyte theory also shows that counterion charge density $\zeta=Ze/V$ ($V$ is the volume occupied by the ion) should control RNA stability. 
As $\zeta$ increases, RNA stability should increase -- a prediction that was validated using a combination of PE-based simulations and experiments. For 
instance, it was shown that the changes in stability of \emph{Tetrahymena} ribozyme in various Group II metal ions (Mg$^{2+}$, Ca$^{2+}$, Ba$^{2+}$, and Sr$^{2+}$) showed a remarkable linear variation with $\zeta$ \cite{Koculi07JACS,Lambert07JMB}. 
The extent of stability is greatest for ions with largest $\zeta$ (smallest $V$ corresponding to Mg$^{2+}$). 
We further showed using Brownian dynamics simulations  that this effect could be captured solely by non-specific ion-RNA interactions \cite{Koculi07JACS}. 
These findings and similar variations of stability in different sized diamines show that (i) the bulk of the stability arises from non-specific association of ions with RNA, and (ii) stability can be greatly altered by valence, shape, and size of the counterions.

\subsection*{Statistics of base pairs in RNA structures}
Because of the systematic experimental studies pioneered by Tinoco, Uhlenbeck, Crothers, and P{\"o}rschke in 1970s \cite{Tinoco02Book,Bloomfield00Book,Porschke74BPC} and Turner and Zuker in 1980-90s  \cite{Turner88ARBC,WalterPNAS94,Xia98Biochem} good estimates of 
the free energies of RNA at the level of secondary structure are available. However, it is still difficult to
 accurately predict  RNA secondary structures (especially if the number of nucleotides is big) because a substantial portion of the nucleotides do not participate in the ``canonical" GC, AU, GU type of base-pairing with base stacks, whose free energy of association can be accurately measured. 
To quantify the fraction of nucleotides forming base pairs (bps), we searched through RNA structures for pairs of nucleotides (A with U, 
G with C or U) in which at least one of their possible pairs  of donor-acceptor heavy atoms is within a cut-off distance of 4 \AA.
For example, Watson--Crick (WC) AÐU pair is identified if N6 from A and O4 from U or N1 from A and N3 from U is within 4 \AA\  from each other. The 4 \AA\ cut-off, larger 
than the typical 2.8--3.0 \AA\ in an ideal hydrogen bonded base-pair, is chosen to account for possible non-ideal bonds and  imperfections in the RNA structures due to the X-ray resolution or to the NMR NOESY signals \cite{Dima05JMB}. 
Hence, the list of hydrogen bonded pairs includes all the WC and reverse WC, Hoogsteen and reverse Hoogsteen, wobble and reverse wobble pairs for AÐU, GÐC, and GÐU; thus in our statistics base pairs found in pseudoknots are also included. 
(Parenthetically, using a more liberal definition of base pairs Westhof and coworkers \cite{Stombaugh09NAR} have shown that nearly 70 \% of nucleotides are involved in base pairs; however, they included non-canonical base pairs such as G-A, A-C, C-C, A-A, U-U whose free energy of stabilization is not known or measured  in their analysis. Even with this higher estimate it is clear that a large fraction of nucleotides are not accounted for in canonical bp formation.)   
Based on our criterion for bp formation, we counted the number of the bps $N_{bp}$ for a given structure in the PDB, and compared this number with the length, $L$, of the sequence. 
We found that $N_{bp}$ varies linearly with $L$ as expected, and the linear growth of $N_{bp}$ with $L$ is satisfied with great accuracy (Fig.\ref{fig:basepair}(a)). 
If all the nucleotides are engaged in WC type canonical base-pairing, the expected value for the ratio between $N_{bp}$ and $L$ is 0.5. 
However, the slope is only 0.27, i.e., $N_{bp}=0.27\times L$. 
The ratio $N_{bp}/L= 0.27$ implies that only 54 \% ($f_{bp}=0.54$) of nucleotides in a RNA chain form base-pairing. 
The remaining 46 \%, which is a large percentage of nucleotides, remain as single strands forming bulges, loops, or dangling ends, the free energy of which varies significantly depending on their size.  Even with the more liberal definition used by Westhof it is clear that a large fraction of nucleotides do not participate in WC-type base pairing.

Using the calculated relationship between  $N_{bp}$ versus $L$, the total length of a given RNA structure can be decomposed into the base-paired and unpaired regions. 
Thus, 
\begin{equation}
L=N_{ss}+2\times N_{bp}
\end{equation}
or 
\begin{equation}
f_{bp}=\frac{2\times N_{bp}}{N_{ss}+2\times N_{bp}}=\frac{2\langle n_{bp}\rangle}{\langle n_{ss}\rangle+2\langle n_{bp}\rangle}
\label{eqn:fbp}
\end{equation}
where $f_{bp}(=0.54)$ is the proportion of nucleotides making  base pairs in the structure, $N_{ss}$ is the number of single stranded nucleotides and nucleotides making non-canonical base pairings. 
From the statistics of the length of stem surveyed from ribosomal RNA (Fig.\ref{fig:basepair}(b)),  
the average length of stem is $\langle n_{bp}\rangle \approx 4.1$, and hence from Eq.\ref{eqn:fbp} with $f_{bp}\approx 0.54$ the average length of single strand is 
$\langle n_{ss}\rangle\approx 7.0$ \cite{fang2011JPCB}. 

The free energies associated with C-C, A-A, G-A as well as loops, bulges, internal loops, and internal multiloops are 
difficult to determine experimentally.  In addition, the PE effects are hard to take into account accurately. These considerations explain the difficulties in devising 
reliable models for simulations. Nevertheless, we show here using examples from our studies that the two classes of CG models not only provide insights into a number of 
RNA problems but also make specific testable predictions. \\

\section {CG models for RNA}

Although CG models have been applied to a number of problems involving proteins their use in the context of RNA is relatively 
recent \cite{Hyeon05PNAS,whitford2009nonlocal,feng2011cooperative,Whitford10RNA,Hayes12JACS}. To our knowledge, we were the first to introduce  CG models for RNA and use them to 
predict the outcome of single molecule pulling experiments and RNA folding \cite{Hyeon05PNAS,Hyeon06BJ,HyeonBJ07,Hyeon08PNAS,HyeonMorrison09PNAS,Cho09PNAS,Biyun11JACS}, and 
to describe the effects of molecular crowding on the stability of folded states of RNA \cite{Pincus08JACS,Denesyuk11JACS}. 
We introduced two levels of coarse graining for nucleic acids.  The first genre of models utilized the three site interaction (TIS) model, which has been subsequently used 
in interesting applications to problems in DNA biology. In the original applications, we used only semi-realistic parameters in the TIS force field. In the more recent 
developments, we parametrized the energy function to ensure that experimental melting curves for a number of oligonucleotides are quantitatively 
reproduced \cite{Denesyuk13JPCB}, resulting in transferable TIS force field.   To cope with larger systems including complexes with proteins, we also 
developed the Self-Organized Polymer (SOP) model. 

\subsection*{Three Interaction Site (TIS) Representation of RNA}
In the TIS model \cite{Hyeon05PNAS,Denesyuk11JACS}, each nucleotide is replaced by three spherical beads P, S and B, representing a phosphate, 
a sugar and a base (the center panel in Box.\ref{Box_TIS_SOP}a) located at the center of mass of the chemical groups. 
The energy function in the TIS model, $U_{\rm TIS}$, is given by
\begin{equation}
U_{\rm TIS}=U_{\rm BL}+U_{\rm BA}+U_{\rm EV}+U_{\rm ST}+U_{\rm HB}+U_{\rm EL},
\label{TIS}
\end{equation}
corresponding to bond length and angle constraints, excluded volume repulsions, single strand base stacking, 
inter-strand hydrogen bonding and electrostatic interactions, respectively. Interactions resulting from  chain connectivity such as bond lengths, $\rho$, and angles, 
$\alpha$, are constrained by harmonic potentials, $U_{\rm BL}(\rho)=k_{\rho}(\rho-\rho_0)^2$ and 
$U_{\rm BA}(\alpha)=k_{\alpha}(\alpha-\alpha_0)^2$, where the equilibrium values $\rho_0$ and $\alpha_0$ are obtained by 
coarse-graining an ideal A-form RNA helix. 
The values of $k_{\rho}$ and $k_{\alpha}$ are given in \cite{Denesyuk13JPCB}. 

\subsubsection*{Excluded volume interaction}
Excluded volume interaction between the interacting sites is modeled by a Weeks-Chandler-Andersen (WCA) potential~\cite{Chandler83Science},
\begin{align}
U_{\rm EV}(r)&=\varepsilon_0\left[\left(\frac {D_0}{r}\right)^{12}-2\left(\frac {D_0}{r}\right)^{6} + 1\right],\ r\le D_0, \nonumber \\
U_{\rm EV}(r)&=0,\ r > D_0, 
\label{POT2}
\end{align}
where $\varepsilon_0=1$ kcal/mol and $D_0=3.2$ \r{A} for all interacting sites.

\subsubsection*{Stacking Interactions}
We assign stacking interactions, $U_{\rm ST}$, between two consecutive nucleotides along the chain,
\begin{equation}
U_{\rm ST}=\frac{U_{\rm ST}^0}{1+k_r(r-r_0)^2+k_{\phi}(\phi_1-\phi_{1,0})^2+k_{\phi}(\phi_2-\phi_{2,0})^2},
\label{UST}
\end{equation}
where $r$ (\r{A}) and $\phi_1$, $\phi_2$ (rad) are defined in Box.1 (b). The equilibrium values of $r_0$, $\phi_{1,0}$ and $\phi_{2,0}$ are obtained by 
coarse-graining an A-form RNA helix. 
The parameters $k_r$, $k_{\phi}$ and $U_{\rm ST}^0$ are derived from available thermodynamic measurements of single-stranded and 
double-stranded RNA~\cite{Xia98Biochem,Bloomfield00Book,Dima05JMB} as described in detail in~\cite{Denesyuk13JPCB}.

\subsubsection*{Hydrogen Bond Interactions}
The TIS description of an RNA molecule includes coarse-grained hydrogen bond interactions $U_{\rm HB}$, which mimic the atomistic hydrogen bonds present in the 
molecule's PDB structure. Each hydrogen bond found in the PDB structure is modeled by a coarse-grained interaction potential,
\begin{eqnarray}
U_{\rm HB}=U_{\rm HB}^0\times\left[1+5(r-r_0)^2+1.5(\theta_1-\theta_{1,0})^2+1.5(\theta_2-\theta_{2,0})^2\right.\nonumber\\
\left.+0.15(\psi-\psi_{0})^2+0.15(\psi_1-\psi_{1,0})^2+0.15(\psi_2-\psi_{2,0})^2\right]^{-1},
\label{UHB}
\end{eqnarray}
where the geompetrical definition of $r$ (\r{A}) and $\theta_1$, $\theta_2$, $\psi$, $\psi_1$, $\psi_2$ (rad) is illustrated in Box.1 (c).
For canonical Watson-Crick base pairs, the equilibrium values $r_0$, $\theta_{1,0}$, $\theta_{2,0}$, $\psi_0$, $\psi_{1,0}$ and $\psi_{2,0}$ are 
adopted from the coarse-grained structure of an ideal A-form RNA helix. For all other hydrogen bonds, the equilibrium parameters are obtained 
by coarse-graining the PDB structure of the RNA molecule. 
The parameter $U_{\rm HB}^0$ has been calibrated to reproduce experimental melting curves of
RNA hairpins and pseudoknots (see~\cite{Denesyuk13JPCB} for details).
Although the terms $\theta_{1,0}$, $\theta_{2,0}$, $\psi_0$, $\psi_{1,0}$, $\psi_{2,0}$ in Eq.\ref{UHB} are defined based on those of PDB structure and Eq.\ref{UHB} is for the native hydrogen bonds, it is also possible to consider generic angles and extend $U_{HB}$ for non-native hydrogen bonds interactions. 

\subsubsection*{Electrostatic interactions}
To model electrostatic interactions, we employ the Debye-H{\"u}ckel approximation \cite{Li13JPCB} combined with the concept of counterion 
condensation~\cite{Manning69JCP}. Using this approach, the electrostatic energy of an RNA conformation is computed as
\begin{equation}
U_{\rm EL}=\frac{Q^2e^2}{2\varepsilon}\sum_{i,j}\frac{\exp\left(-|{\bf r}_i-{\bf r}_j|/\lambda\right)}{|{\bf r}_i-{\bf r}_j|},
\label{GDH}
\end{equation}
where $Q$ is the reduced charge of the phosphate groups, $e$ is the proton charge,  $|\mathbf{r}_i-\mathbf{r}_j|$ is the distance between 
two phosphates $i$ and $j$, $\varepsilon$ is the dielectric constant of water and $\lambda$ is the Debye-H{\"u}ckel screening length. In monovalent salt 
solutions, Manning's theory of counterion condensation predicts~\cite{Manning69JCP}
\begin{equation}
Q = \frac{b}{l_{\rm B}},
\label{Manning}
\end{equation}
where $b$ is the length per unit charge in the polyelectrolyte in the absence of counterion condensation and $l_{\rm B}$ is the Bjerrum length, 
$l_{\rm B}=e^2/\varepsilon k_{\rm  B}T$. 
We previously established that using $b=4.4$ \r{A} gives good agreement between the simulation and experiment for the RNA 
thermodynamics in monovalent salt solutions~\cite{Denesyuk13JPCB}.

\subsection*{Self-Organized Polymer (SOP) Model}
The self-organized polymer (SOP) model, a new class of versatile coarse-grained structure-based models, is well suited to understanding dynamics at the spatial resolution that single-molecule spectroscopy of large RNA provide. 
We refer to the model as the SOP model because it only uses the polymeric nature of the biomolecules and the crucial topological constraints that arise from the specific fold. 
We  introduced the SOP model to study the response of proteins and RNA to mechanical force. 
The reason for using the SOP model in force spectroscopy applications is the following: 
(i) Forced-unfolding and force-quench refolding lead to large conformational changes on the order 10--100 nm. 
Currently, single molecule experiments (laser optical tweezers or atomic force microscopy) cannot easily resolve structural changes below 1 nm. As a result, details of the rupture of hydrogen bonds or local contacts between specific residues cannot be discerned from force-extension curves or the dynamics of the end-to-end distance $R$ alone. 
Because only large changes in $R$ are monitored, it is not crucial to model details due to local interactions such as bond-angle and various dihedral angle potentials. 
(ii) In the context of mechanical unfolding as well as the folding of RNAs, many of the details of the unfolding and folding pathways can be accurately computed by taking into account only the interactions that stabilize the native fold. 
Previous studies also suggested that it is crucial to take into account chain connectivity and attractive interactions that faithfully reproduce the contact map of a fold. The basic idea of the SOP model is to use the simplest possible Hamiltonian to simulate the low-resolution global dynamics for biopolymers of arbitrary size. 
The potential function for biopolymers in the SOP representation is 
\begin{align}
H_{SOP}=&-\sum_{i=1}^{N-1}\frac{k}{2}R_0^2\log{\left[1-\frac{\left(r_{i,i+1}-r_{i,i+1}^0\right)^2}{R_0^2}\right]}
+\sum_{i=1}^{N-2}\epsilon_l\left(\frac{\sigma}{r_{i,i+2}}\right)^6
\nonumber\\
&+\sum_{i=1}^{N-3}\sum_{j=i+3}^N\epsilon_h\left[\left(\frac{r_{i,j}^0}{r_{i,j}}\right)^{12}-2\left(\frac{r_{i,j}^0}{r_{i,j}}\right)^6\right]\Delta_{ij}+\sum_{i=1}^{N-3}\sum_{j=i+3}^N\epsilon_l\left(\frac{\sigma}{r_{i,j}}\right)^6(1-\Delta_{ij})
\label{eqn:SOP}
\end{align}
The first term in Eq.\ref{eqn:SOP} is the finite extensible nonlinear elastic (FENE) potential for chain connectivity with parameters, $k = 20$ kcal/(mol \AA$^2$), $R_0=0.2$ nm. 
The distance between neighboring beads at $i$ and $i+1$ is $r_{i,i+1}$,  
and $r^0_{i,i+1}$ is the distance in the native structure. 
The use of the FENE potential is more advantageous than the standard harmonic potential, especially for
the purpose of simulating force-induced stretching of biopolymer, because the fluctuations of $r_{i, i + 1}$ are strictly restricted around $r_{i,i+1}^0$ with variations of $\pm R_0$, which enable us to produce worm-like chain behavior without including additional bond-angle potential term to the energy potential. 
The second term ensures non-crossing of the chain by making $i$, $i + 2$ pairs interact repulsively with $\sigma=3.8$ \AA. 
The Lennard-Jones potential with a special condition for the native pairs is used to account for interactions that stabilize the native fold. 
The native pairs are defined for beads, between $i$ and $j$ with $|i-j|>2$, whose distance in the native state is less than 8 \AA. 
We use $\epsilon_h=$1--2 kcal/mol for native pairs, and $\epsilon_l=1$ kcal/mol for nonnative pairs. 
Although potential attraction between nonnative pairs are neglected in the SOP model, this should not qualitatively affect the results for forced unfolding 
of RNA. Especially when simulating force-induced unfolding dynamics of native structure, such interactions do not make significant contribution to the dynamics. 
To probe forced unfolding of RNA, it is sufficient to only include attractive interactions between contacts that stabilize the native state. 

There are five parameters in the SOP force field. In principle, the ratio of $\epsilon_h$ and $\epsilon_l$ can be adjusted to obtain agreement with experiments. 
For simplicity, a uniform value of $\epsilon_h$ is assumed for proteins or RNA. But, if such modification is critical for understanding the dynamic behavior of the molecular system of interest, $\epsilon_h$ can be made sequence-dependent and ion-implicit.
The time spent in calculating the Lennard-Jones potential scales as $\sim\mathcal{O}(N^2)$.
From algorithmic point of view, drastic savings in computational time can be achieved by truncating forces due to
the Lennard-Jones potential for interacting pairs with $r_{ij} > 3r_{ij}^0$ or 3$\sigma$ to zero, and by taking into account the neighboring list for numerical enumeration.  Variation of the original SOP model has been used with remarkable success to predict the outcomes of single molecule pulling experiments. These predictions were subsequently validated.

\section{Applications}

\subsection*{Complex kinetics of hairpin formation}
When viewed on length scales that span several base pairs and time scales on the order of $\mu$s  folding of a RNA hairpin is simple. Hairpin formation can be described using a two-state model. 
However, when probed on short times (ns--$\mu$s range) the formation of a small hairpin, involving a loop and base-stacking, is  complex. 
Temperature jump experiments have shown  that the kinetics of hairpin formation in RNA deviates from the classical two-state kinetics, and is best described as a multi-step process \cite{Ma06JACS}. 
Additional facets of hairpin formation have been revealed in single molecule experiments that use mechanical force ($f$). Inspired by these experiments we performed simulations using the TIS model, with energy function give in~\cite{Hyeon05PNAS}, by varying $T$ and $f$~\cite{Hyeon08JACS}.
The equilibrium phase diagram showed two basins of attraction (folded and unfolded) at the locus of critical points $(T_m, f_m)$, which are the transition midpoints separating the unstructured and hairpin states.  At $T_m$ and $f_m$ the probability of being unfolded and folded is the same.
The free energy surface obtained from simulations  explained the sharp bimodal transition between the folded and unfolded state when the RNA hairpin is subject 
to $f$ \cite{Hyeon05PNAS,Hyeon08JACS}, characteristic of a two-state system. 

However, upon temperature quench, hairpin forms by multiple steps \cite{Hyeon08JACS}  as observed in the kinetic experiments~\cite{Ma06JACS}. 
Folding pathways between $T$-quench and $f$-quench refolding are markedly different (Fig.\ref{fig:RNA_hairpin}b), reflecting the differing initial conditions. 
The initial conformations generated by forced unfolding are fully extended and structurally homogeneous.
The first event in folding, upon $f$-quench,  is loop formation, which is a slow nucleation process (see Fig.\ref{fig:RNA_hairpin}b). 
Zipping of the remaining base pairs leads to rapid hairpin formation.
Refolding upon $T$-quench commences from a structurally broad ensemble of unfolded conformations. 
Therefore, nucleation can originate from many regions in the molecule (see Fig.\ref{fig:RNA_hairpin}b). 
Our simulations showed that the complexity of the folding kinetics, observed in ribozyme experiments \cite{Pan97JMB}, is  reflected in the formation of simple RNA hairpin \cite{Chen00PNAS,Hyeon05PNAS}.

\subsection*{Mechanical unfolding of RNA}
The stability of RNA molecule in the native state can be roughly (neglecting entropy) decomposed into the secondary and tertiary interactions,  
$\epsilon^{tot}=\sum_i\epsilon_i^{sec}+\sum_k\epsilon_k^{ter}$, where $i$ and $k$ refer to the number of secondary and tertiary structural elements, respectively. 
The difference in the free energy between secondary and tertiary interactions $\sum_i\epsilon_i^{sec}\gg \sum_k\epsilon_k^{ter}$ ensures that force disrupts tertiary 
contacts prior to secondary structures; thus, rips in force-extension curves (FECs) are due to disruptions of domains defined by secondary structural elements. The separation in the energy scales ($\sum_i\epsilon_i^{sec}\gg \sum_k\epsilon_k^{ter}$) enables analysis of FEC for RNA to be made domain by domain. The hairpin stacks, which can vary in the length and sequence, are the simplest structural motifs, and the presence of hairpin loops, bulges, internal loops, and internal multiloops increase the complexity of RNA structures. 
The effect of force on RNA structures with increasing complexity from simple hairpins, hairpins with bulge and internal loop to three-way junction is quantitatively 
described elsewhere \cite{HyeonBJ07}. We showed that many aspects of the physics of mechanical unfolding of RNA, such as force-dependent free energy profile as a function 
of  the end-to-end distance, loading rate dependent movement of transition state, and pulling speed dependent unfolding pathways, can be 
dissected using structural motifs.\\

\subsubsection*{The L-21 \emph{T}. ribozyme.}
The \emph{Tetrahymena} ribozyme and its independently folding subdomains (P4-P6 and P5abc) have been extensively studied since the discovery of self-splicing enzymatic 
activity in 1980s \cite{HerschlagJMB03,ThirumARPC01,Treiber01COSB}. 
By probing the unfolding characteristics of increasingly larger constructs of the \emph{T}. ribozyme by using LOT experiments, Onoa et al. \cite{Bustamante03Science} were 
able to associate the force peaks in the FECs to rupture of specific substructures, providing an outline of the forced unfolding pathway of RNA. 
They assumed that extension by a certain length corresponds to unraveling of the entire helical substructures, and inferred the unfolding pathway of ribozymes from FECs 
alone. However, it is difficult to unambiguously assign the specific paired helices for a given FEC. The number of rips in the FECs also varies depending on the specific 
molecule that is being stretched. In addition, there are multiple unfolding routes \cite{Bustamante03Science}, indicative of heterogeneity in force-induced unfolding.

To provide molecular details of forced unfolding, we used the SOP representation of the  Westhof model (TtLSU.pdb from the website http://www-ibmc.u-strasbg.fr/upr9002/westhof.) \cite{WesthofCB96}, and computed the FEC  \cite{WesthofCB96} using  Brownian dynamics simulations. 
In agreement with LOT experiments \cite{Bustamante03Science}, in the majority of the cases 
the FECs have about eight peaks. The number of peaks varies from molecule to molecule, which is increasingly being recognized as a characteristic of heterogeneity at the single-molecule level \cite{Hyeon2012NatureChem,ZhuangSCI02,Solomatin10Nature}. 
Explicit comparison of the FECs with the dynamics of the contact disruption (Fig.\ref{fig:T_ribozyme}b) enables us to read off the molecular events associated with each rip in the FEC. 
There are two major unfolding pathways (Fig.\ref{fig:T_ribozyme}). 
(i) [N] $\rightarrow$ [P9.2] $\rightarrow$ [P9.1, P9, P9.1a] $\rightarrow$ [P2] $\rightarrow$ [P2.1] $\rightarrow$ [P3, P7, P8] $\rightarrow$ [P6] $\rightarrow$ [P4, P5] $\rightarrow$ [P5a, P5b, P5c]. 
(ii) 
[N] $\rightarrow$ [P2] $\rightarrow$ [P2.1] $\rightarrow$ [P9.2] $\rightarrow$ [P9, P9.1, P9.1a] $\rightarrow$ [P3, P7, P8] $\rightarrow$ [P6] $\rightarrow$ [P4, P5] $\rightarrow$ [P5a, P5b, P5c]. 
The difference between the two pathways is the switch in the order of unfolding of the peripheral domains (P2 and P9). 
The experimentally inferred, most probable pathway is [N] $\rightarrow$ [P9.2] $\rightarrow$ [P9.1] $\rightarrow$ [P9, P9.1a] $\rightarrow$ [P2, P2.1] $\rightarrow$ [P3, P7, P8] $\rightarrow$ [P6, P4] $\rightarrow$ [P5] $\rightarrow$ [P5a, P5b, P5c]. 

Snapshots from simulations in Fig.\ref{fig:T_ribozyme}c show the conformational changes that occur in the force-induced unfolding transition. 
Although the predicted FECs do not quantitatively agree with the measurements due to the differences between the loading rates and the spring constant used in the simulations and experiments,  the order of unfolding of the helices and the heterogeneous nature of the unfolding pathways are consistent with experiments.
A few additional comments about our results from SOP simulations are worth making. 
\begin{enumerate}
\item Both simulations and experiments \cite{Bustamante03Science} find that the peripheral domains unravel before disruption of the tertiary interactions involving the catalytic core. Complete rupture occurs when helices P6, P4, and P5abc unfold. 
\item The major and minor unfolding pathways are not trivially related to each other. 
In one pathway, unfolding starts from P2, while in the other unraveling starts from the P9 end. 
From a structural perspective, P2 forms tertiary interactions with P5c, whereas the P9 helix is in contact with P5. 
The free energies of the tertiary interactions involving the P2 and P9 domains are also different. 
Thus, from both the energetic and structural considerations, the differences in the unfolding pathways are significant. 
\item The rips corresponding to the peripheral domains P9 in the simulations are [P9.2] $\rightarrow$ [P9.1, P9, P9.1a], whereas in the experiments three rips corresponding to [P9.2] $\rightarrow$ [P9.1] $\rightarrow$ [P9, P9.1a] are identified. 
The two rips corresponding to [P2] $\rightarrow$ [P2.1] also differ from the single rip [P2, P2.1] in the experiment. 
The minor differences may be due to the slight variations in the constructs used in experiments (390-nt) versus simulations (407-nt). 
\end{enumerate}

\subsubsection*{Stretching \emph{Azoarcus} Ribozyme.} 
Building on the good agreement between unfolding pathways in simulations and single molecule experiments on {\it Tetrahymena} ribozyme, we performed simulations for a smaller but structurally related ribozyme.  Mechanical unfolding trajectories of the 195-nt \emph{Azoarcus} ribozyme (PDB code: 1u6b), generated at three different loading rates using the SOP model, reveal distinct unfolding pathways (Fig.\ref{fig:Azoarcus}). 
At the highest loading rate, the FEC has six conspicuous rips~\cite{Hyeon06Structure}, whereas at the lower loading rate the number of peaks is reduced to between two and four. These simulations showed that the unfolding pathways can be altered by changing the loading rate.  

At the highest loading rate, the dominant unfolding pathway is  N $\rightarrow$ [P5] $\rightarrow$ [P6] $\rightarrow$ [P2] $\rightarrow$ [P4] $\rightarrow$ [P3] $\rightarrow$ [P1]. 
At medium loading rates, the ribozyme unfolds via N $\rightarrow$ [P1, P5, P6] $\rightarrow$ [P2] $\rightarrow$ [P4] $\rightarrow$ [P3], which produces four rips in the FECs. 
Multiple helices in the square bracket mean that they unravel nearly simultaneously. 
At the lowest loading rate, the ribozyme unfolds by two steps via N $\rightarrow$ [P1, P2, P5, P6] $\rightarrow$ [P3, P4]. 
The simulations using the SOP model indicate that unfolding pathways depend on the loading rate, a result that does not seem to be well appreciated. The physical origin of the change in the unfolding pathways as the loading rate is varied is explained elsewhere \cite{Hyeon06Structure}.

These results showed that the unfolding pathways can drastically change depending on the loading rate, $r_f$. 
The dominant unfolding rate depends on $r_f$, suggesting that the outcomes of unfolding by LOT and AFM experiments can be dramatically different. Our predictions for the response of {\it Azoarcus} ribozyme can be tested readily using pulling experiments. \\

\subsection*{Towards folding under cellular conditions} 

\subsubsection*{Background} 
Because the cytosol is crowded, replete with macromolecules such as ribosomes, lipids, proteins, and RNA, whose collective  volume fraction ($\phi$) can exceed 0.2, it is expected that RNA folding in the crowded environment is different from 
{\it in vitro} experiments that are  conducted under infinite dilution conditions \cite{Elcock10COSB,Cheung05PNAS,Dhar10PNAS,Homouz08PNAS}. 
Describing the transition between folded and unfolded states of RNA under 
crowded conditions is complicated because the nature of the interactions between the crowding agents and RNA  is not fully understood. However, to a first approximation, the dominant effect of 
crowding agents is to exclude the molecule of interest from the volume occupied by crowders (see Box.\ref{DepletionForce} for depletion interaction arising from excluded volume interactions). If excluded volume interactions 
dominate (an assumption that has to be tested before the theory can be applied to analyze experiments), then the stability 
of the folded state of the RNA is enhanced, compared to the infinite dilution limit.  In this case, the loss 
in entropy of the folded state due to crowding is much less than that of the unfolded state, resulting in the 
stabilization of the folded state. The resulting entropic stabilization of the folded state~\cite{Cheung05PNAS,Minton05BJ}  has been recently affirmed  in a number of studies~\cite{Pincus08JACS,Kilburn10JACS,Denesyuk11JACS}. 

We showed, using theoretical arguments and coarse-grained simulations, that crowding can modestly stabilize RNA secondary structures~\cite{Pincus08JACS}. However, RNA requires counterions 
(Mg$^{2+}$, for example) for tertiary folding. Thus, the effect of macromolecular crowding on tertiary structures of RNA 
may be complicated depending on the interplay between electrostatic and excluded volume interactions. Using small angle X-ray scattering measurements it has been shown that, in the presence of 
polyethylene glycol (PEG), the 195 nucleotide {\it Azoarcus} ribozyme is more compact relative 
to $\phi = 0$~\cite{Kilburn10JACS}. It was concluded that excluded volume effects play a dominant role in the 
compaction of RNA in low molecular weight PEG. Interestingly, the transition to the folded state occurred at a lower 
Mg$^{2+}$ concentration in the presence of PEG~\cite{Kilburn10JACS}. Even if excluded volume interactions largely 
determine the stability of the folded states of RNA, a number of variables besides $\phi$, such as size and shape 
of crowding agents, also contribute to the stability of RNA in the presence of inert crowding agents. Thus, 
a systematic study of the influence of macromolecular crowding on RNA is required. To date there are only a handful of works that have considered this problem.

\subsubsection*{Crowding shifts conformational equilibrium between pseudoknot (PK) and hairpin (HP) in human telomerase RNA} 
 
 As a biologically relevant example, we 
simulated crowding effects on the transition between the hairpin (HP) and pseudoknot (PK) conformations (Fig.\ref{fig:PK_HP}) in the 
pseudoknot domain of human telomerase RNA (hTR)~\cite{Theimer03PNAS}. 
The activity of the pseudoknot domain, which is conserved in different organisms, is closely linked to chromosome stability~\cite{Blasco03COGD,Chen04PNAS}. 
Mutations that either increase or decrease the stability of the PK conformation result in a reduction in telomerase 
activity~\cite{Comolli02PNAS,Theimer05MolCell}. 
Therefore, it is important to compare the impact of physical factors, such as macromolecular crowding, with that of naturally occurring chemical mutations. 

To provide quantitative estimates of  crowding-induced changes in the stability of RNA,  we augmented the TIS model for RNA with interactions between the crowders and the nucleotides. 
We modified the  Lennard-Jones potential  to model interactions of 
RNA with the spherical crowders of arbitrary size,
\begin{eqnarray}
U_{\rm LJ}(r)&=&\varepsilon\frac{2R_i}{D_0}\left[\left(\frac {D_0}{r + D_0 - D}\right)^{12} 
- 2\left(\frac {D_0}{r + D_0 - D}\right)^{6} + 1\right],\ r\le D, \nonumber \\
U_{\rm LJ}(r)&=&0,\ r > D, 
\label{POT1}
\end{eqnarray}
where $r$ is the distance between the centers of mass of two interacting particles, $D_0$ is the effective penetration 
depth of the interaction, $R_i$ is the radius of an RNA coarse-grained bead, $r_{\rm C}$ is the radius of a crowder,
and $D=R_i+r_{\rm C}$. The ratio $2R_i/D_0$ in Eq. (\ref{POT1}) is used to scale the interaction strength 
$\varepsilon$ in proportion to the surface contact area. This potential accounts for nonspecific 
surface interactions between spherical crowders representing large macromolecules and individual segments of the 
coarse-grained RNA. The global effects of hard sphere crowders on RNA can be qualitatively explained using the concept of depletion forces, first discussed by Asakura and Oosawa (see Box 3). 

\subsubsection*{Crowding effect is negligible for large crowders}

The high density of macromolecules in the cell (volume fractions $\phi \approx 0.2-0.4$) reduces the space available for 
conformational fluctuations. Therefore, macromolecular crowding should shift the thermodynamic equilibrium 
between the HP and PK states of the hTR pseudoknot domain towards the more compact PK. To assess the extent to which
PK is favored at $\phi \ne  0$, we first discuss the simulation results~\cite{Denesyuk11JACS} for the HP and PK states of the 
modified pseudoknot domain, $\Delta$U177 (Fig.\ref{fig:PK_HP}). The molecular construct $\Delta$U177 has been examined experimentally 
{\it in vitro} at $\phi$ = 0~\cite{Theimer05MolCell}.  The atomistic structures of the HP and PK conformations of $\Delta$U177 
are available from the Protein Data Bank, codes 1NA2 and 2K96, respectively.

Here,  we only consider spherical crowders with radius 
$r_{\rm C}$. For monodisperse particles, the volume fraction is $\phi = 4 \pi r_{\rm C}^3 \rho/3$, where $\rho$ is 
the number density. Thus, $\phi$ can be changed by increasing or decreasing $\rho$ or by altering the size of the crowding 
particles. For clarity of presentation  we fix $\phi = 0.3$ and examine the consequences of changing $r_{\rm C}$. Based on general theoretical 
considerations~\cite{Asakura58JPS,Shaw91PRA} (see Box 3) it can be shown that, in the colloid limit $r_{\rm C}> R_{\rm G}^0$, the 
crowding agents would have negligible effects on RNA stability. Here, $R_{\rm G}^0$ is the size of RNA in the absence of 
the crowding agent. It is only in the opposite polymer limit, $r_{\rm C}< R_{\rm G}^0$, that the crowding particles 
would affect RNA stability. We therefore expect that the magnitude of the crowding effect should depend only on the ratio 
$r_{\rm C}/R_{\rm G}^0$. Although this scaling type result follows from theory it has not been adequately tested in experiments involving RNA.

The HP melting profile, taken to be the negative derivate of the number of intact base pairs 
$N_{\rm BP}$ with respect to $k_{\rm B} T$, is plotted in Fig.\ref{fig:rC}a for the crowder radius $r_{\rm C}=26$ \r{A}~\cite{Denesyuk11JACS}. 
Such crowders are larger than the radius of gyration of strand G93--C121 in the unfolded state, $R_{\rm G}^0=20$ \r{A}. As discussed above, large crowders should have minimal effects on the melting of the HP even at $\phi=0.3$. For a 
fixed $\phi$, the average distance between two spherical crowders will increase with the crowder size. If the unfolded 
hairpin can easily fit in the interstitial space, the folding/unfolding transition will not be affected significantly by 
the presence of crowders. For $\phi=0.3$ and the crowder radius $r_{\rm C}=26$ \r{A}, which is only slightly larger than 
$R_{\rm G}^0$, the increase $\Delta T$ in the melting temperature is 1.5 $^{\circ}$C for stem 1 of the HP and is 
negligible for stem 2 (Fig.\ref{fig:rC}a). Further increase in $r_{\rm C}$ results in $\Delta T\approx0$ for both stems (data not 
shown).

Fig.\ref{fig:rC}a also shows the melting profile of the HP in a ternary mixture of crowders, containing volume fractions 
$\phi=0.11$, 0.11 and 0.08 of particles with $r_{\rm C}=104$ \r{A}, 52 \r{A} and 26 \r{A}, respectively~\cite{Denesyuk11JACS}. The sizes and volume fractions of individual components in the model mixture correspond 
to the ribosome, large enzymatic complexes and relatively small individual proteins, found in {\it E. coli}. Because all 
the values of $r_{\rm C}$ in the {\it E. coli} mixture are larger than $R_{\rm G}^0$, we expect only small changes in the 
melting profile of the HP (Fig.\ref{fig:rC}a). For the total volume fraction of 0.3, the melting temperature of the HP stem 1 
increases only by 2 $^{\circ}$C with respect to $\phi=0$ (Fig.\ref{fig:rC}a). Interestingly, the effect of the {\it E. coli} 
mixture is similar in magnitude to that of a monodisperse suspension with $r_{\rm C}=26$ \r{A} and $\phi=0.3$. In contrast,
a monodisperse suspension with $r_{\rm C}=26$ \r{A} and $\phi=0.08$, which is equivalent to the smallest particle component 
in the mixture, has negligible effect on the melting of the HP (Fig.\ref{fig:rC}a). 

In summary, the crowding effect of polydisperse mixtures is largely the effect of the smallest particle component, 
but taken at the total volume fraction of the mixture. The excess stability of the 
folded state due to crowding decreases nonlinearly with the radius of the crowding particle $r_{\rm C}$ (see the next subsection). We therefore 
propose that, for crowding in the cellular environment, the main role of large macromolecules is to increase the 
effective volume fraction of the relatively small macromolecules. 

\subsubsection*{Role of crowder size in the PK-HP equilibrium}

The change in stability of the HP and PK at 37 $^{\circ}$C in the presence of  monodisperse crowders for different 
crowder radii $r_{\rm C}$ ($\phi=0.3$) shows that the magnitude of the excess 
stability $\Delta G(0.3)-\Delta G(0)$ is small if $r_{\rm C}/R_{\rm G}^0>1$, and increases sharply for 
$r_{\rm C}/R_{\rm G}^0<1$. Crowding effect is larger for the PK for all values of $r_{\rm C}$ (Fig.\ref{fig:rC}b), 
indicating an equilibrium shift towards this conformation.  Macromolecular crowding promotes the binding of strand C166--A184 to the remainder of the structure in the PK 
(Fig.\ref{fig:PK_HP}), because fluctuations associated with this strand are restricted in a crowded environment. The crowder radius 
$r_{\rm C}=12$ \r{A} corresponds to the size of an average protein {\it in vivo}. For $\phi=0.3$ and 
$r_{\rm C}=12$ \r{A}, $\Delta G_{\rm PK}(0.3)-\Delta G_{\rm PK}(0)=-2.4$ kcal/mol and 
$\Delta G_{\rm HP}(0.3)-\Delta G_{\rm HP}(0)=-1.0$ kcal/mol, which results in the relative stabilization of the PK 
conformation by $-1.4$ kcal/mol (Fig.\ref{fig:rC}b). 

\subsubsection*{Implications for function}

It is known that changes in the relative stability of the HP and PK conformations of the hTR pseudoknot 
domain compromise the enzyme activity. The estimate of the crowding effect in a typical cellular environment,
$\Delta\Delta G=-1.4$ kcal/mol (see above), allows us to assess the extent to which macromolecules could regulate telomerase 
activity. We showed~\cite{Denesyuk11JACS} that the enzyme activity of the hTR mutants~\cite{Comolli02PNAS,Theimer05MolCell} decreases exponentially 
as a function of $|\Delta\Delta G^*|=|\Delta G^*_{\rm PK}(0)-\Delta G_{\rm PK}(0)|$, 
where $\Delta G^*_{\rm PK}(0)$ and $\Delta G_{\rm PK}(0)$ are the stabilities of mutant and wild-type pseudoknots at 
$\phi=0$ (Fig.\ref{fig:activity}). The majority of mutations destabilize the PK, $\Delta\Delta G^*>0$ (black squares in Fig.\ref{fig:activity}) and only two 
mutants have $\Delta\Delta G^*<0$ (red stars in Fig.\ref{fig:activity}). For destabilizing mutants the reduction in activity, 
$\alpha$, was shown to follow the exponential dependence~\cite{Denesyuk11JACS}, $\alpha=\exp (-0.6\Delta\Delta G^*)$ 
(thick curve in Fig.\ref{fig:activity}). The naturally occurring destabilizing mutations DKC and C116U have been linked to diseases 
dyskeratosis congenita and aplastic anemia, respectively~\cite{Vulliamy02Lancet,Fogarty03Lancet}. The DKC and the 
stabilizing $\Delta$U177 mutations have been studied {\it in vivo} (green symbols in Fig.\ref{fig:activity}), as well as 
{\it in vitro}. In both cases, mutant telomerase {\it in vivo} was found to be significantly less active than the 
corresponding construct {\it in vitro}, suggesting that a number of factors determine the activity of telomerase 
{\it in vivo}. 

Although macromolecular crowding enhances the stability of the PK state, the crowding effect ($\Delta\Delta G=-1.4$ kcal/mol) is less than 
the stability changes caused by mutations. In Fig.\ref{fig:activity} the grey area marks the domain of potential mutants with 
$\Delta\Delta G^*>0$, whose activity may be completely restored by macromolecular crowding. All experimentally studied 
mutants fall outside the marked domain, including the two disease related mutants DKC and C116U. Nevertheless, due to the 
strong dependence of enzyme activity on $\Delta\Delta G^*$, the effect of crowding on telomerase function may be 
significant. We estimate that the activity of telomerase can be up- or down-regulated by more than two-fold in response 
to density fluctuations in its immediate environment. Furthermore, due to the expected dynamical heterogeneities in cells, 
there will be variations in enzyme activity in different cell regions. It should be emphasized that the enhancement in the stability of PK 
relative to HP is a maximum for a given $r_{\rm C}$ and $\phi$ in the excluded volume dominated regime. All specific interactions between crowders 
and nucleotides will invariably decrease the extent of stability changes. Thus, the predicted enhancement is an upper bound.

\subsubsection*{Crowding effects on RNA at different ionic strengths}

Entropic stabilization mechanism \cite{Cheung05PNAS} implies that crowding increases the stability of the folded state by reducing the
population of expanded conformations in the unfolded state. Therefore, we expect that the magnitude of the crowding 
effect will be sensitive to the ionic strength of the RNA buffer, since the latter determines the size of the 
unfolded RNA. The quantitative discussion above assumed the limit of high ionic strength. As the buffer ionic concentration $c$ is lowered, the screening of the negative charge on the RNA 
sugar-phosphate backbone becomes less efficient, which in turn increases the mean radius of gyration of conformations in 
the unfolded state. The function $R_{\rm G}(c)$ for the unfolded PK is shown in Fig.\ref{fig:Salt}a in the absence (black diamonds) 
and presence (green circles) of crowding. The same general trend is observed in both cases, with the $R_{\rm G}$ values 
being consistently smaller when crowders are present for the entire range of $c$. In accord with our predictions, the 
crowder-induced stabilization of the folded PK becomes more significant at low ionic strengths (red squares in Fig.\ref{fig:Salt}a). 

Interestingly, the stabilization effect increases rapidly upon lowering $c$ from 1 M to 0.1 M, but shows little change
when $c$ is lowered further to below 0.1 M. The underlying reason for such behavior can be traced to the probability 
distributions $p(R_{\rm G})$ in the unfolded state (Fig.\ref{fig:Salt}b). For a given crowder solution, we can identify a typical 
size of the cavity which will be free of any 
crowders. If the radius of gyration of RNA conformations is such that they fit into the cavity, these conformations will 
not be perturbed by crowding. On the other hand, the population of conformations with $R_{\rm G}$ larger than the typical 
cavity size will be significantly depleted by crowding. For $\phi=0.3$ and $r_{\rm C}=12$ \r{A}, we can infer the size of 
a standard empty cavity from the distributions $p(R_{\rm G})$ at high ionic strength (Fig.\ref{fig:Salt}b). Note that $p(R_{\rm G})$ 
decreases for $R_{\rm G}>20$ \r{A} when crowders are present (green solid line in Fig.\ref{fig:Salt}b), but $p(R_{\rm G})$ increases 
with $R_{\rm G}$ around 20 \r{A} in the absence of crowders (black solid line in Fig.\ref{fig:Salt}b). This indicates that the 
crowders significantly perturb the unfolded conformations with $R_{\rm G}$ larger than 20 \r{A}, which can serve as an 
upper estimate of the smallest RNA size affected by crowders. When $c$ decreases, the distribution $p(R_{\rm G})$ shifts to 
larger $R_{\rm G}$, increasing the fraction of the unfolded conformations affected by crowders. At $c=0.1$ M, all 
statistically significant values of $R_{\rm G}$ in the unfolded state fall within the range $R_{\rm G}>20$ \r{A}, so that
the entire distribution $p(R_{\rm G})$ is depleted due to crowding (symbols in Fig.{\ref{fig:Salt}b). 
This explains why the 
crowding-induced stabilization is almost constant below 0.1 M, even if the average $R_{\rm G}$ continues to increase 
rapidly all the way to 0.01 M (symbols in Fig.\ref{fig:Salt}a). This explanation also shows that the relative ratio between the size of the unfolded 
RNA and the crowder size dictates the extent of stability changes. 

\subsubsection*{RNA becomes compact as $\phi$ increases}

The reduction in conformational space accessible to RNA should increase with $\phi$, for a fixed $r_{\rm C}$. Thus, we 
expect that $R_{\rm G}$ should decrease as $\phi$ increases. This is precisely what is observed in experiments, which show 
that at all concentrations of Mg$^{2+}$ the {\it Azoarcus} ribozyme becomes more compact as the volume fraction of the 
crowding agent (PEG) increases~\cite{Kilburn10JACS} (Fig. 9). Interestingly, the midpoint of the folding 
transition $c_{\rm m}$ -- the concentration of Mg$^{2+}$ at which the folded and unfolded states of the {\it Azoarcus} 
ribozyme have equal populations -- also decreases as $\phi$ increases (Fig. 9). This finding can be readily 
explained in terms of the entropic stabilization mechanism~\cite{Cheung05PNAS} and suggests that, to a first approximation,
PEG behaves as an inert hard sphere crowding agent. Based on our considerations from the previous section, we predict that 
the shift in $c_{\rm m}$ due to crowding will also depend on the concentration of monovalent counterions in the RNA buffer,
a prediction that is amenable to experimental test. In addition, it would be of interest to perform experiments at a fixed 
$\phi$ but varying $r_{\rm C}$, which can be changed by decreasing or increasing the molecular weight of PEG.

\section{Outlook}

There are compelling reasons to develop CG models further in order to study many complex phenomena associated with RNA. The two
most obvious ones are: (i) Current atomistic RNA force fields do not appear to be accurate enough to predict folding 
thermodynamics of even a tetra loop \cite{Garcia13PNAS}. It is unclear if the deficiency is in the parametrization of RNA or water models or both. 
(ii) The time scales associated with the processes that we have touched upon here are far too great for even the most specialized computers.  
By following essentially methodology of calibrating energy functions by direct comparison with experiments we have shown that CG models can not 
only reproduce experimental measurements but also can make testable predictions. Although not discussed here,  the folding landscape of {\it add}-riboswitch 
(activates translation in {\it E. Coli.} by binding to adenine) under force was predicted accurately \cite{Lin08JACS} three years before experimental validation. 
More recently, we have predicted that a single point mutation should alter the folding landscape of {\it add}-riboswitch to resemble that of {\it pbuE}-riboswitch, 
which activates transcription \cite{Lin14PCCP}. These results show that discovery of design principles of riboswitches, which control gene expression, 
can be  made using CG models but remain, at present, outside the scope of atomically detailed simulations. The concept of modeling at the appropriate level, prevalent 
in the study of synthetic materials, which also applies to biological problems, will surely spur us on to develop suitable CG models and theories that capture the
essence of the problem at hand without being encumbered by unnecessary details, and will continue to grow because there is an appetite to understand the workings of a cell.   \\

{\bf Acknowledgements:} This work was supported by a grant from the National Science Foundation (CHE09-10433) (D.T.). The list of references is not exhaustive and should be viewed as a guide to the literature.
\\

\clearpage

\renewcommand\figurename{Box.} 
\begin{wrapfigure}{r}{0.5\textwidth}
\includegraphics[width=3.5in]{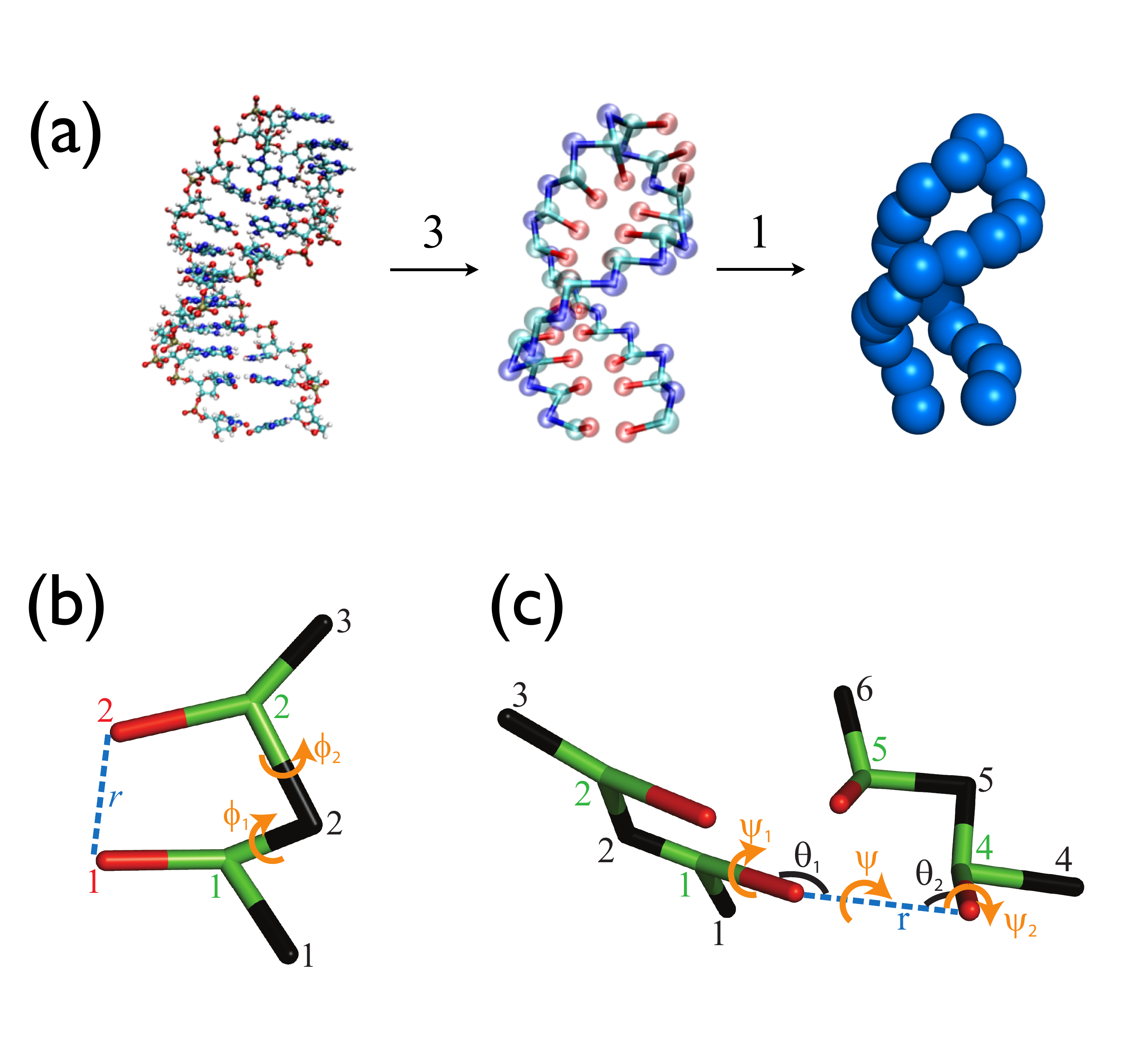}
\caption{
\label{Box_TIS_SOP}}
\end{wrapfigure}
Box \ref{Box_TIS_SOP}: {\bf Genre of CG models.}

In response to the challenge of describing biological processes that span several orders of magnitude in time and length scales a variety of CG models for DNA, RNA and proteins have been proposed. 
Although CG models have been prevalent in the polymer literature for over fifty years their use in proteins began in earnest with the pioneering work of Levitt and Warshel \cite{Levitt75Nature}.  The efficacy of off-lattice models for protein folding kinetics was first demonstrated by Honeycutt and Thirumalai \cite{Honeycutt90PNAS}.  More recently, in the same spirit we introduced CG models for RNA.
In all the CG models  polypeptide chains and nucleic acids are represented using a reduced description.  
The accompanying figures show
{\bf a}. Three Interaction Site (TIS) models for a RNA hairpin obtained by representing each nucleotide by three sites corresponding to phosphate, ribose 
and base (the center panel).  The rightmost panel shows further reduction to the SOP model. The associated energy functions are described 
in the main text. 
{\bf b}. Definition of the distance between the bases ($r$) and of the two dihedral angles ($\phi_1$, $\phi_2$) for stacking interactions in the TIS model 
(Eq.~\ref{UST}). Phosphates, sugars and bases are shown in black, green and red, respectively. The indices refer to different nucleotides.
{\bf c}. Definition of the distance ($r$), angles ($\theta_1$, $\theta_2$) and dihedral angles ($\psi$, $\psi_1$, $\psi_2$) for hydrogen bond interactions between
the bases in the TIS model (Eq.~\ref{UHB}). Definitions of the structural parameters for hydrogen bonds involving phosphates or sugars and further details on the
energy functions can be found in \cite{Denesyuk13JPCB}.

A major advantage of CG models is that their conformational space can be exhaustively sampled. 
However, even with simplification accurate results for thermodynamics might require enhanced sampling methods. 
Towards this end simulation of CG models have used replica exchange methods and multicanonical methods. In addition, low friction Langevin dynamics has also been used to efficiently sample conformational space. These methods are necessary especially in simulating proteins with complex topology.  In order to obtain kinetic information for folding  typically Brownian dynamics (BD) simulations are performed. 
In BD simulations the Brownian time is $\tau_H\approx \zeta_Ha^2/k_BT_s$ where $\zeta_H$ is the friction constant, $a$ is roughly the size of a 
coarse-grained bead, and $T_s$ is the simulation temperature. Estimate of these quantities \cite{VeitshansFoldDes97} have been used to map simulation 
times to real times in a number of applications. \\

\renewcommand\figurename{Box.} 
\begin{wrapfigure}{r}{0.6\textwidth}
\includegraphics[width=4.5in]{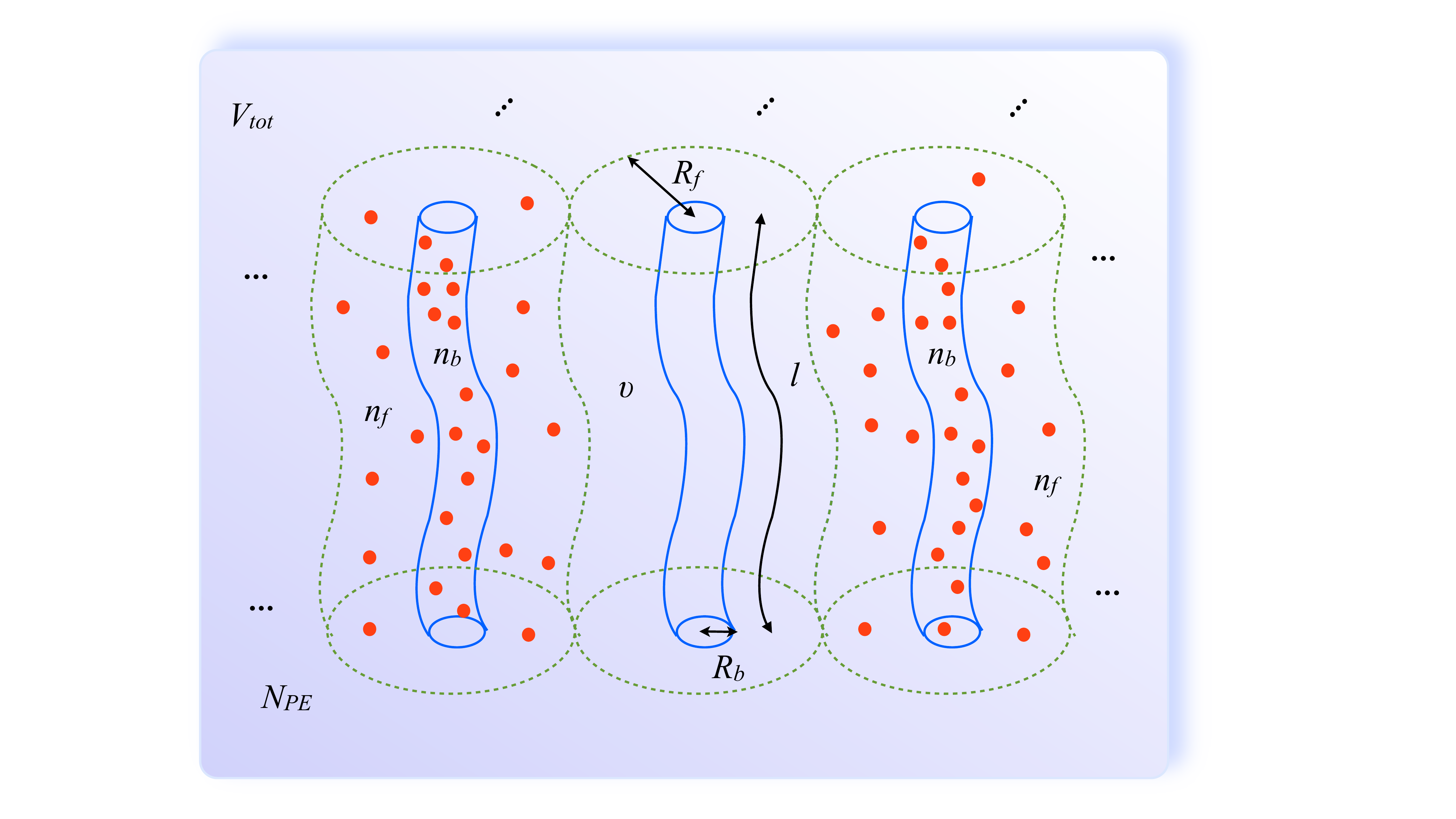}
\caption{\label{PE}}
\end{wrapfigure}
Box \ref{PE}: {\bf Manning-Oosawa mechanism.} 
Counterion condensation onto a polyelectrolyte (PE) is analogous to vapor-to-liquid phase transition \cite{OosawaBook}. 
Consider a system consisting of $N_{PE}$ polyelectrolyte molecules in a solution of volume $V$. 
Each PE, whose contour length is $l$, has $n$ charged groups on the backbone onto which $n$ monovalent counterions can condense. 
Counterions are divided into $n_f$ free (vapor) and $n_b$ bound (liquid) counterions. 
Figure shows that a volume occupied by the $n_b$ bound counterions around each PE is  approximated as a cylinder with thickness $R_b$ and volume $v$ ($\pi R_b^2 l\approx v$). 
Average volume for $n$ counterions released from each PE is $\pi R_f^2l\approx V/N_{PE}$.
Therefore, the volume fraction of PE ($\phi\equiv N_{PE}v/V$) is related to $R_b$ and $R_f$ as $(R_f/R_b)^2=\phi$. 
At equilibrium, chemical potentials of free ($\mu_f$) and bound ions ($\mu_b$) are equal, 
$\Delta \mu = \mu_b-\mu_f=0$.     
The energetic contribution $\Delta \epsilon$ results from the difference in the Coulomb potentials ($-e\psi$) between the two phases. 
For a PE modeled as a cylinder $\Delta \epsilon=-e(\psi_b-\psi_f)=-e(\psi(R_b)-\psi(R_f))=2k_BT\beta\xi\log{(R_b/R_f)}\approx k_BT\beta\xi\log{\phi}$ where 
$\beta\equiv n_f/n$ and $\xi=nl_B/l$, $l_B=e^2/k_BT$ is the Bjerrum length. 
The entropic contribution $T\Delta s$ to $\Delta \mu$ is determined by the ratio of accessible volumes for the counterions in each phase, $V_b$ for bound 
ions and $V_f$ for free ions.  Since $V_b\approx N_{PE}v$ and $V_f=V-N_{PE}v$, $T\Delta s=k_BT\log\left(V_b/V_f\right)=k_BT\log{\frac{\phi}{1-\phi}}$. 
For $\phi\rightarrow 0$ 
$\Delta \mu\approx (\beta\xi-1)k_BT\log{\phi}$. This suggests that the fraction of free ions at equilibrium ($\Delta \mu=0$) is $\beta=1/\xi$ 
and the fraction of ions bound to PE (or the fraction of backbone charge neutralized by the counterions) is $1-\xi^{-1}$. 
One can use similar arguments to quantify the effect of counterion condensation onto RNA, which is roughly spherical in the folded 
state \cite{Hyeon06JCP_2}. Surprisingly, this concept familiar in PE theory has been very successful in estimating renormalized charges on the phosphate groups.\\

\renewcommand\figurename{Box.} 
\begin{wrapfigure}{r}{0.5\textwidth}
\includegraphics[width=3.5in]{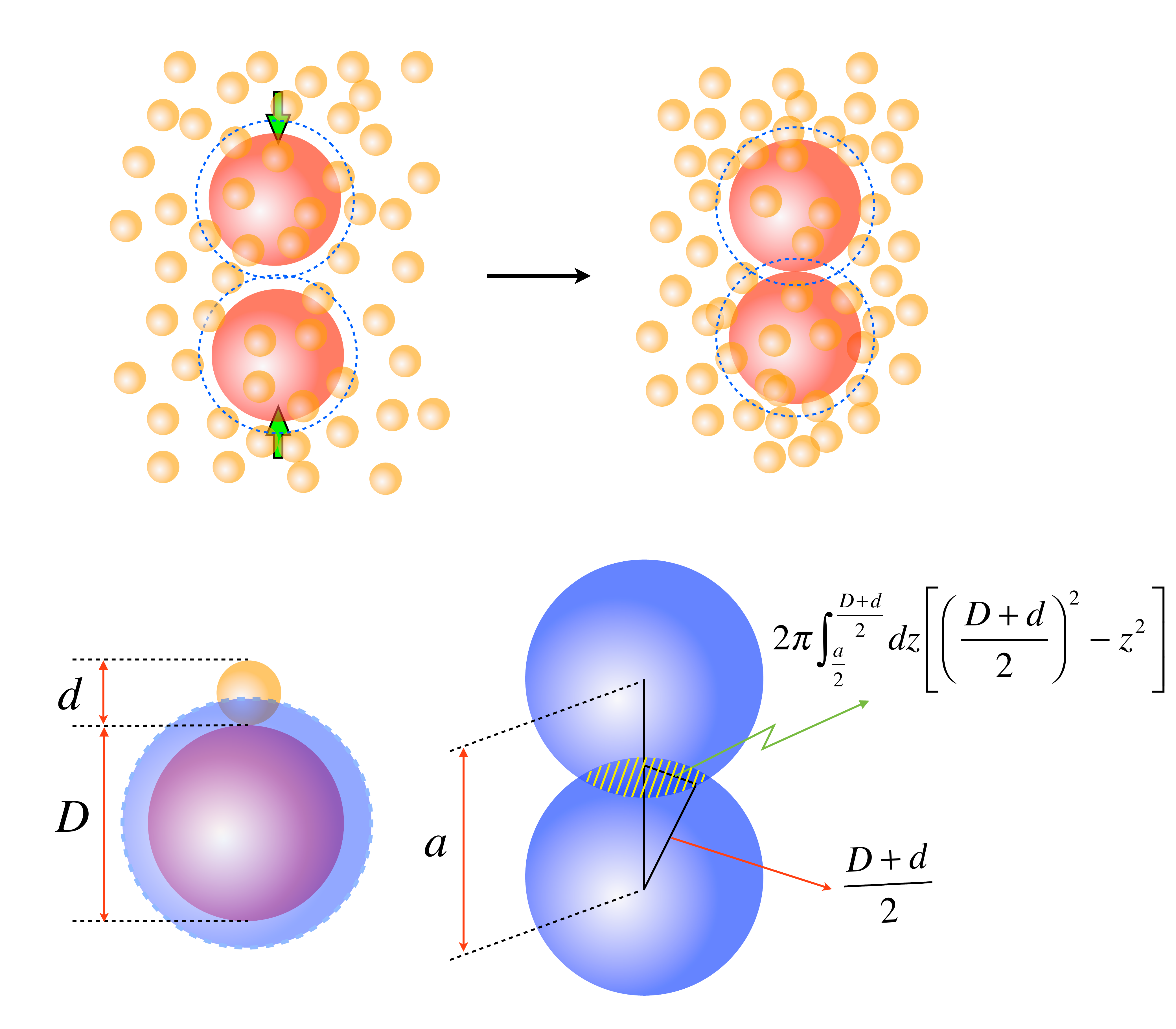}
\caption{ \label{DepletionForce}}
\end{wrapfigure}
Box \ref{DepletionForce}: {\bf Depletion force.}
If the spacing between two colloidal particles in a suspension containing inert polymers is less than the polymer size, osmotic pressure of the solution produces an attractive force, referred to as  \emph{depletion force}, between the colloidal particles \cite{Asakura58JPS}.  
Depletion force arises so as to maximize the entropy of macromolecules.  
Consider two colloidal particles (red spheres with diameter $D$) suspended in a solution of volume $V$ with macromolecules (orange spheres with diameter $d$) ($D>d$). 
If both colloids and macromolecules are hard spheres,  
the partition function for the system when two colloidal particles are separated by a distance $a$ is the accessible volume for macromolecules, i.e., 
$Q(a)=\int_V dx \exp{\left(-\omega(x,a)\right)}=V_{eff}(a)$, 
where $\omega(x,a)$ is the potential energy between the two colloidal particles at separation $a$ and $x$ denotes the orientation $(\theta,\phi)$ between them.   
Depending on the distance between the two colloidal particles $a$, the accessible volume for the macromolecules is 
\begin{align}
V_{eff}(a)=\left\{ \begin{array}{ll}
     V-\pi\left[\frac{4}{3}\left(\frac{D+d}{2}\right)^3+\left(\frac{D+d}{2}\right)^2a-\frac{1}{12}a^3\right]& \mbox{$(D\leq a\leq D+d)$}\\
     V-\frac{8\pi}{3}\left(\frac{D+d}{2}\right)^3& \mbox{$(a\geq D+d)$}\end{array}\right.
\label{eqn:fluctuation}
\end{align}
For $D\leq a\leq D+d$ the two spheres with radius $\frac{D+d}{2}$ overlap, reducing the accessible volume for macromolecules. Given the partition function, one can 
calculate the depletion force between the spheres using $f=Nk_BT\partial\log Q(a)/\partial a$. For $V\rightarrow \infty$, 
\begin{align}
f_{eff}(a)=\left\{ \begin{array}{ll}
     -p\times \frac{\pi}{4}\left[(D+d)^2-a^2\right]& \mbox{$(D\leq a\leq D+d)$}\\
     0& \mbox{$(a\geq D+d)$}\end{array}\right.
     \label{eqn:depletion}
\end{align}
Here $p=Nk_BT/V$ is the osmotic pressure due to the macromolecules. 
Note that $\frac{\pi}{4}\left[(D+d)^2-a^2\right]$ is the area of circular intersection formed by the overlap of two spheres with 
$\sqrt{\left(\frac{D+d}{2}\right)^2-\left(\frac{a}{2}\right)^2}$ being the radius of intersection. 
Attractive interaction between two colloids (green arrows in the top figure) due to depletion force manifest itself only if the separation of the 
colloidal particles ($a-D$) is less than the size of macromolecules ($d$). 
From Eq.\ref{eqn:depletion}, it is expected that for a given osmotic pressure ($p$) the depletion force will be larger for bigger-size macromolecular crowders.
For a given volume fraction of crowders $\varphi_c=\frac{4}{3}\pi(d/2)^3N/V$, the osmotic pressure is inversely proportional to $d^3$ ($p=6\varphi_c k_BT/\pi d^3$); thus 
depletion force becomes stronger for crowders with a smaller size $d$ although the range of non-zero depletion force $[a\in (D,D+d)]$ is smaller. 
\\

\clearpage

\section*{Figure Captions}

{\bf Fig.\ref{fig:basepair}. Statistics of base pairs in RNAs.}
{\bf a.} The dependence of the number of base-pairs in a given native structure as a function of sequence length, $L$. The figure was adapted from Dima \emph{et al.} (2005) \cite{Dima05JMB}. 
{\bf b.} Distribution of the number of consecutive base pairs composing stem, calculated from 16S, 23S \emph{E}. coli and \emph{T}. thermophilus ribosomes.  
\\

{\bf Fig.\ref{fig:RNA_hairpin}. Folding of RNA hairpin.}
{\bf a.} Kinetics of hairpin formation under $f$-quench and $T$-quench conditions. 
In each molecule the hairpin formation time can be decomposed into looping and zipping time as $\tau_F= \tau_{loop} + \tau_{zip}$. 
The dynamics of looping is more time-consuming in $f$-quench condition than in $T$-quench condition 
(compare $P^f_{loop}(t)$ and $P^T_{loop}(t)$), but the dynamics of zipping occurs on a similar time scale (compare $P^f_{zip}(t)$ and $P^T_{zip}(t)$).
The probability of molecules remaining unfolded at time $t$ ($P_U(t)$) is plotted in the inset. 
Under $f$-quench and $T$-quench conditions the probabilities are fit to $P_U^f(t) = e^{-(t -50)/138}$ for $t > 50$ $\mu$s and $P_U^T(t)=0.44 e^{-t/63} + 0.56 e^{-t/104}$, respectively, where the time $t$ is in the unit of $\mu$s. 
In particular, the lag phase of $P_U^f(t)$ at $0 < t < 50$ $\mu$s suggests an obligatory intermediate ($I^f_{SL}$ where $SL$ stands for small loops). 
{\bf b.} Schematic of refolding pathways of a RNA hairpin upon quenching the force from a high to low value (left) and obtained from temperature quench (right) using TIS model. 
Upon $f$-quench folding commences from an extended ($E$) state by forming the turn, which nucleated the hairpin formation (left). However, folding occurs by multiple pathways upon $T$-quench. 
The figures were adapted from Hyeon and Thirumalai (2008) \cite{Hyeon08JACS} (reproduced with copyright permission from the American Chemical Society). 
\\

{\bf Fig.\ref{fig:T_ribozyme}. Unfolding of \emph{T}. ribozyme.}
{\bf a.}
The dynamics of disruption of individual contacts ($Q_i(t)$) for one of the trajectories in the major unfolding pathway. The scale in differing shades represents the number of contacts that survive at $t$. 
The tertiary contacts indicated with the circles (a--d) are shown in the \emph{T}. ribozyme structure. The yellow squares show interactions that stabilize the P3 pseudoknot. 
The figure on right shows that unfolding can also occur by an alternate pathway in which the initial event is the opening of P2. 
{\bf b.} Snapshots of \emph{T}. ribozyme structures are shown along the major and minor pathways.
The figures adapted from Hyeon \emph{et al.} (2006) \cite{Hyeon06Structure} (reproduced with copyright permission from Elsevier). 
 \\

{\bf Fig.\ref{fig:Azoarcus}. Loading rate dependent variation of unfolding pathway of \emph{Azoarcus} ribozyme}. The secondary structure of the ribozyme is shown on the left.
Time evolutions of contact disruption at three different loading rates $r_f=1.2\times 10^3$ pN/s, $3.6\times 10^2$ pN/s, and $1.9\times 10^1$ pN/s are shown from the top to bottom.
The figure adapted from Hyeon \emph{et al.} (2006) \cite{Hyeon06Structure} (reproduced with copyright permission from Elsevier). 
  \\
  
{\bf Fig.\ref{fig:PK_HP}. Secondary and tertiary structures of the pseudoknot (PK) and hairpin (HP) conformations for $\Delta$U177. }
{\bf a.} PK secondary and tertiary structures, {\bf b}. HP secondary and tertiary structures. Nuclear magnetic resonance (NMR) structure of the HP includes residues G93 to 
C166 only (PDB code 1NA2). To quantify the effect of crowders on the PK--HP equilibrium, we added an unstructured tail A167--A184 to the NMR structure in {\bf a}. 
This inclusion ensures that identical RNA sequences are used in simulations of the PK and HP conformations. \\

{\bf Fig.\ref{fig:rC}.}
{\bf a.} Melting profiles of the HP in various crowding environments at $\geq 1$ M monovalent salt concentration. Black solid curve : without crowders.
Red dashed curve (the model Escherichia coli mixture) : $\phi= 0.11$ of crowders with $r_{\rm C} = 104$ \r{A}, $\phi = 0.11$ of crowders 
with $r_{\rm C} = 52$ \r{A} and $\phi = 0.08$ of crowders with $r_{\rm C} = 26$ \r{A}. Green dashed-dotted curve : 
$\phi = 0.08$ of crowders with $r_{\rm C} = 26$ \r{A}. Blue dotted curve : $\phi = 0.3$ of crowders with 
$r_{\rm C} = 26$ \r{A}. Thick orange curve is the experimental UV data at 200 mM KCl from Fig. 2b in 
Theimer et al. (2003) \cite{Theimer03PNAS}, divided by $5.53\times 10^{-5}$. The two peaks indicate melting of stems 1 and 2 of the HP. 
The peak positions (melting temperatures) and the overall width of the melting curve (melting range) serve as a measure of agreement between theory and spectroscopic data. 
{\bf b}. Changes in stability (kcal/mol) of the HP and PK at 37 $^{\circ}$C due to crowders at $\phi=0.3$, as a function of the crowder 
radius $r_{\rm C}$. 
The figure adapted from Denesyuk and Thirumalai (2013) \cite{denesyuk2013BiophysRev} (reproduced with copyright permission from Springer).
\\
  
{\bf Fig.\ref{fig:activity}.}
Activity of mutant telomerase normalized to wild-type activity (100 \%) as a function of the magnitude of the stability difference between mutant and wild-type pseudoknots.
Three mutants, $\Delta$U177, DKC, and C116U, are explicitly marked. The in vitro data for destabilizing (black squares) and stabilizing (red stars) mutations are from 
Table 2 in Theimer et al. (2005) \cite{Theimer05MolCell}. Green symbols : in vivo data for $\Delta$U177 and DKC from Comolli et al. (2002) \cite{Comolli02PNAS}. 
The experimental data for destabilizing mutations is fit to the exponential function (solid curve). Gray area : range of stability differences that can be 
accommodated by crowding. The figure adapted from Denesyuk and Thirumalai (2011) \cite{Denesyuk11JACS} (reproduced with copyright permission from the American Chemical Society). \\

{\bf Fig.\ref{fig:Salt}.} 
{\bf a.} The radius of gyration $R_{\rm G}$ of the unfolded PK (right axis) as a function of the monovalent ion concentration $c$ for $\phi = 0$ (black 
diamonds) and $\phi = 0.3$, $r_C$ = 12 \r{A} (green circles). Red squares show the excess stability of the PK due to crowding, 
$\Delta G_{\rm PK}(0.3)-\Delta G_{\rm PK}(0)$, for $r_{\rm C}=12$ \r{A} (left axis). 
{\bf b.} Probability distributions $p(R_{\rm G})$ of the radius of gyration of the unfolded PK for $\phi=0$ and $c=1$ M (black solid line), 
for $\phi=0$ and $c=0.1$ M (black diamonds), for $\phi=0.3$, $r_{\rm C}=12$ \r{A} and $c=1$ M (green solid line), and for $\phi=0.3$, 
$r_{\rm C}=12$ \r{A} and $c=0.1$ M (green circles). Vertical dashed line indicates the smallest size of RNA conformations that will be perturbed by 
crowders with $\phi = 0.3$ and $r_{\rm C} = 12$ \r{A}. \\

{\bf Fig.\ref{fig:Rg_Mg}.}
Small angle X-ray scattering measurements of the radius of gyration of the \emph{Azoarcus} ribozyme for different concentrations of Mg$^{2+}$ ions and 
polyethylene glycol (PEG). Graphic adopted from Kilburn et al. \cite{Kilburn10JACS}\\

\clearpage 
\setcounter{figure}{0}
\renewcommand\figurename{Fig.} 
\begin{figure}[h]
  \includegraphics[width=\columnwidth]{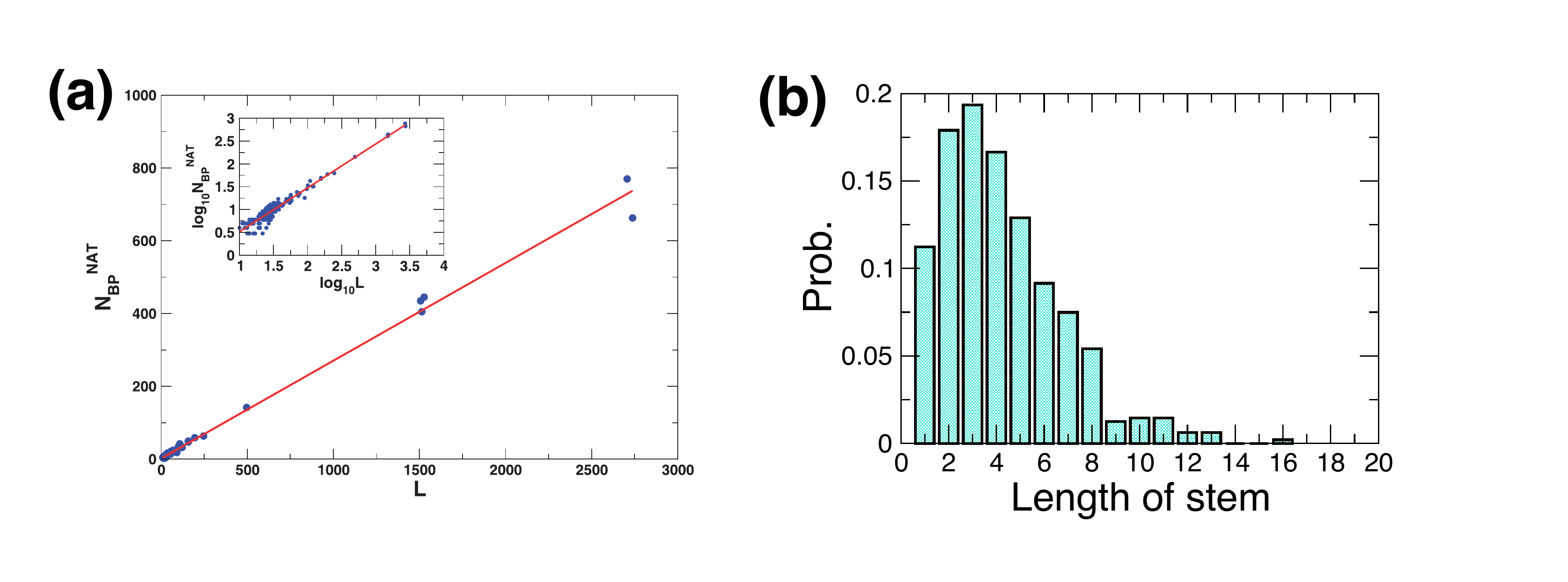}
  \caption{}\label{fig:basepair}
\end{figure}

\begin{figure}[h]
  \includegraphics[width=\columnwidth]{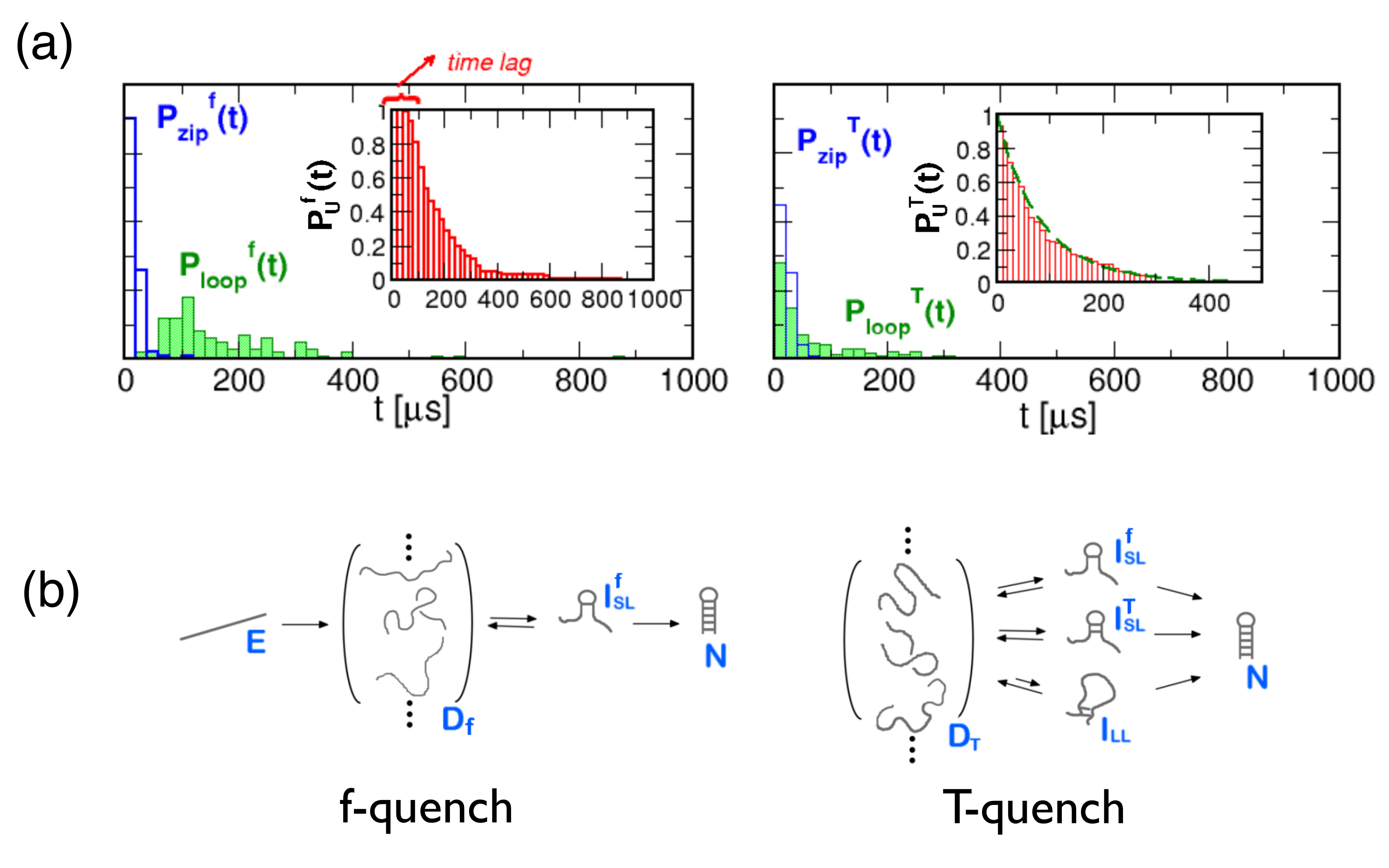}
  \caption{}\label{fig:RNA_hairpin}
\end{figure}
\clearpage 

\begin{figure}[h]
  \includegraphics[width=0.9\columnwidth]{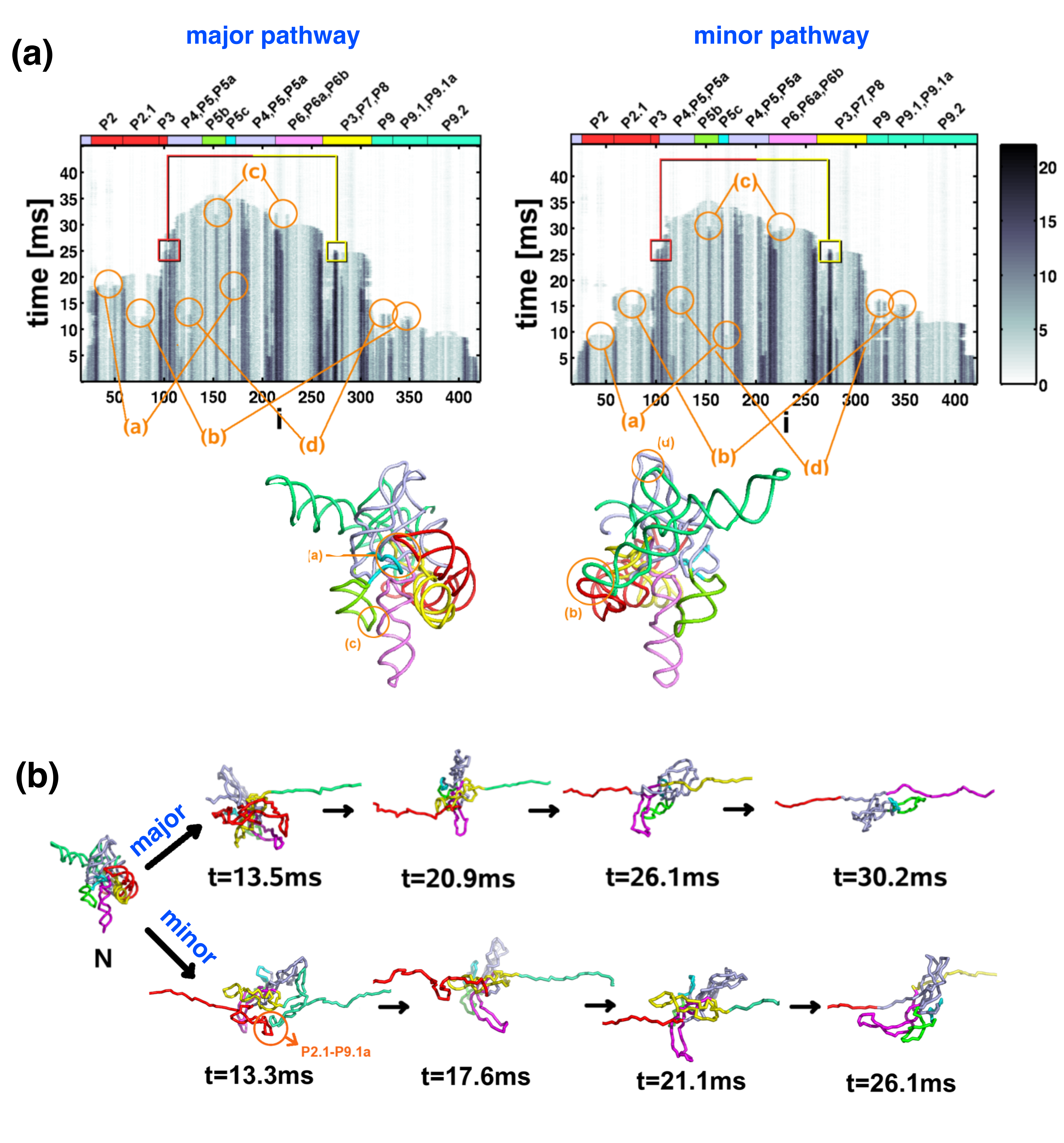}
  \caption{}\label{fig:T_ribozyme}
\end{figure}
\clearpage 

\begin{figure}[h]
  \includegraphics[width=0.9\columnwidth]{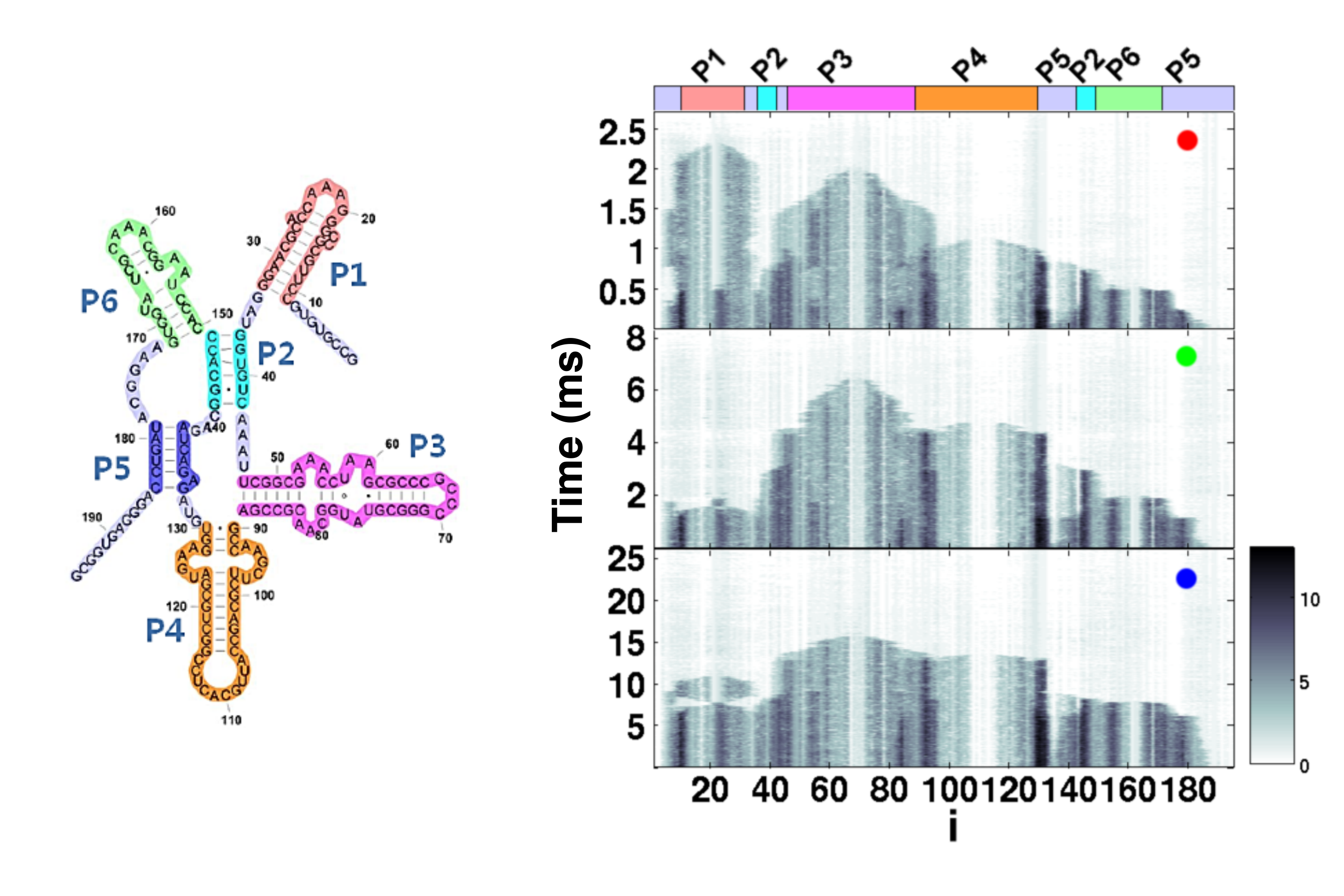}
  \caption{}\label{fig:Azoarcus}
\end{figure}
\clearpage 

\begin{figure}[h]
  \includegraphics[width=0.9\columnwidth]{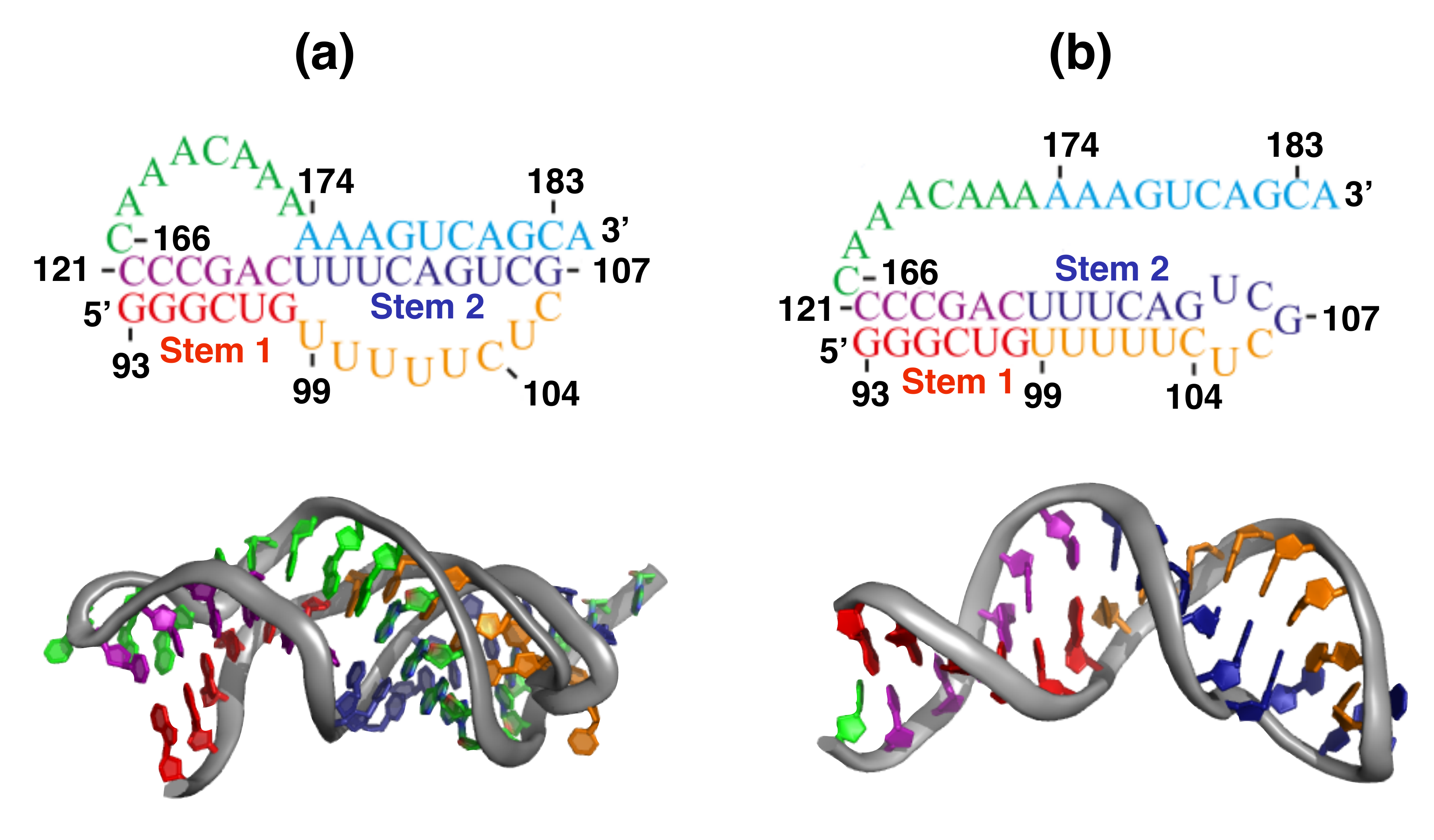}
  \caption{}\label{fig:PK_HP}
\end{figure}
\clearpage 

\begin{figure}[h]
  \includegraphics[width=0.9\columnwidth]{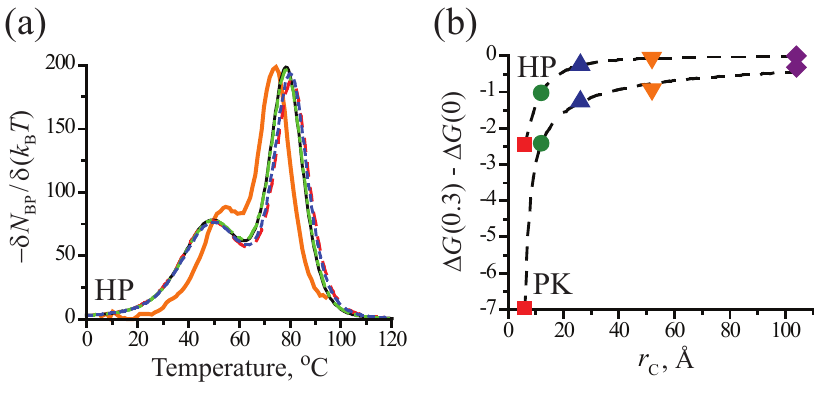}
  \caption{}\label{fig:rC}
\end{figure}
\clearpage

\begin{figure}[h]
  \includegraphics[width=0.8\columnwidth]{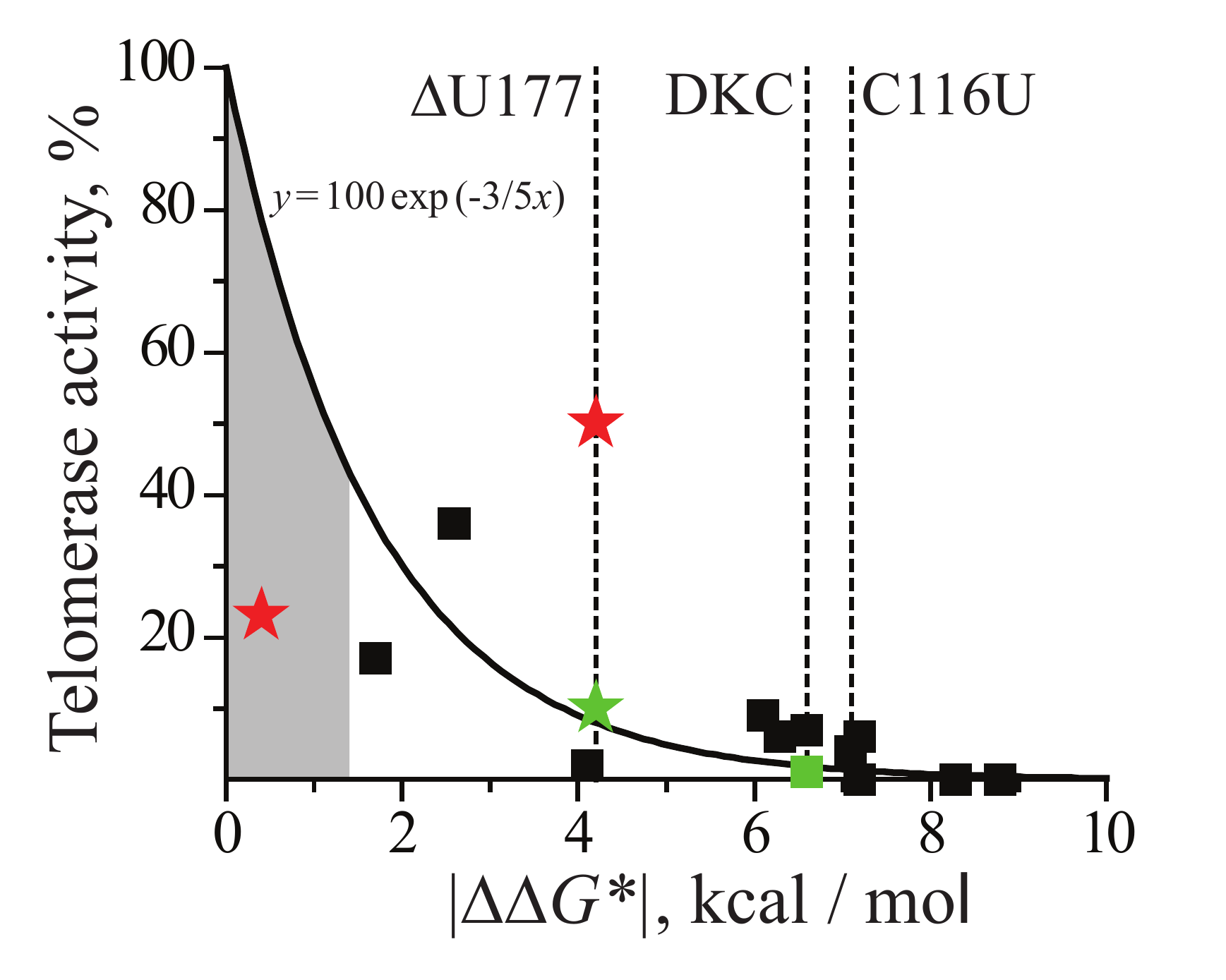}
  \caption{}\label{fig:activity}
\end{figure}
\clearpage

\begin{figure}[h]
  \includegraphics[width=0.8\columnwidth]{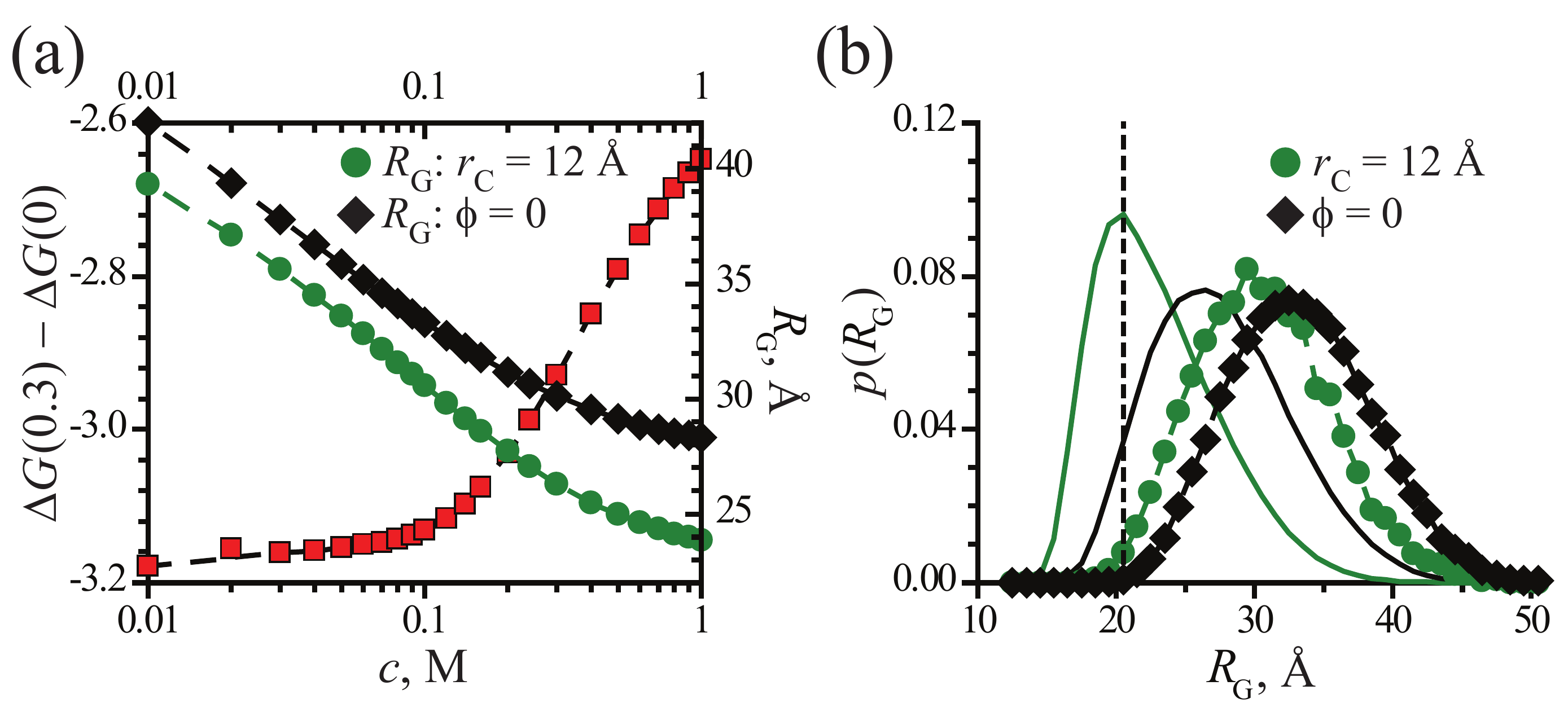}
  \caption{}\label{fig:Salt}
\end{figure}
\clearpage

\begin{figure}[h]
  \includegraphics[width=0.8\columnwidth]{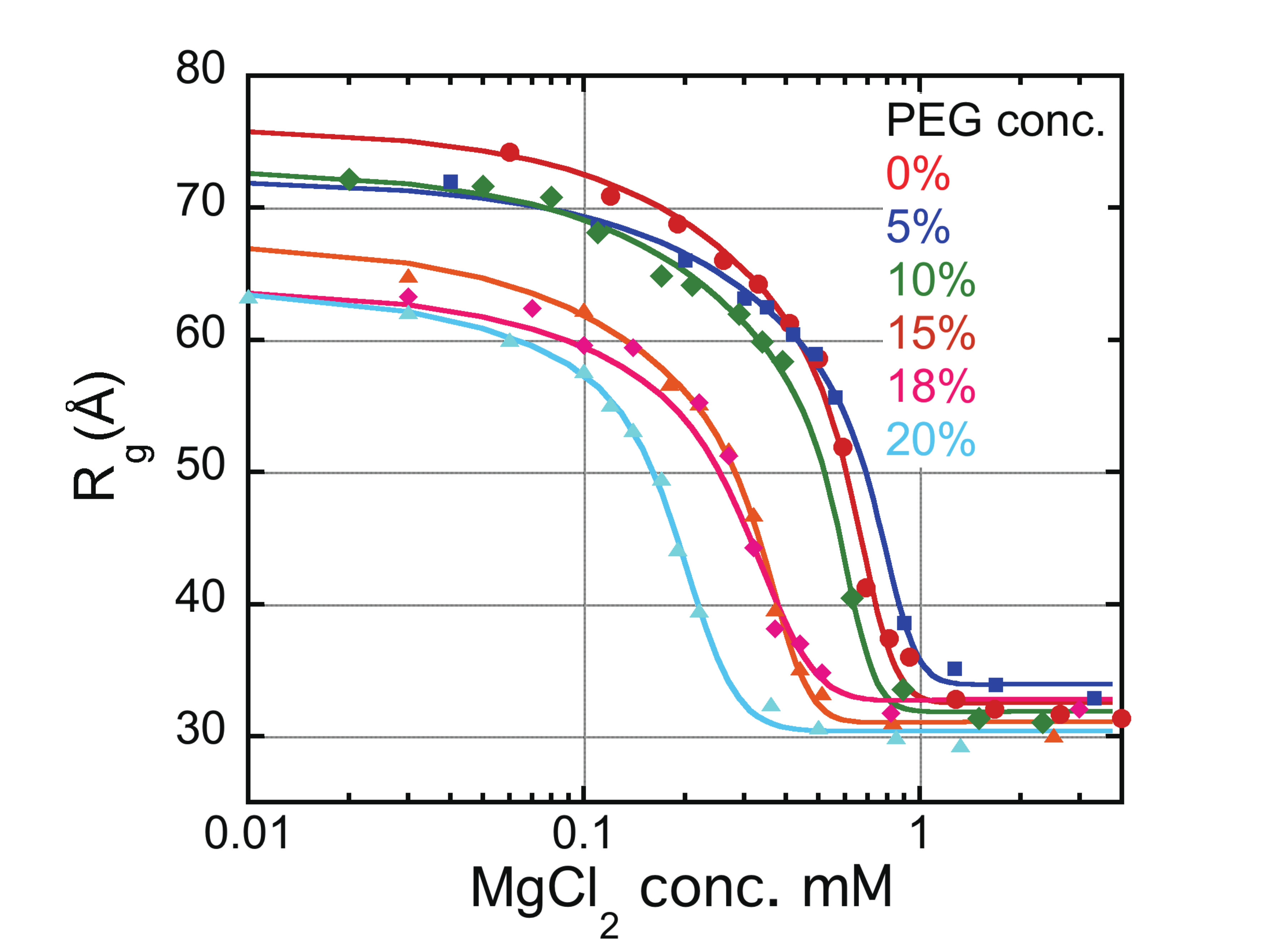}
  \caption{}\label{fig:Rg_Mg}
\end{figure}
\clearpage

\end{document}